\documentclass{article}

\usepackage{arxiv}

\usepackage[utf8]{inputenc} 
\usepackage[T1]{fontenc}    

\usepackage{amsmath}
\usepackage{amssymb}
\usepackage{amsfonts}

\usepackage[numbers]{natbib}

\usepackage{graphicx}
\usepackage{float}
\usepackage{pgfplots}
\pgfplotsset{compat=1.18}

\usepackage{booktabs}
\usepackage{multirow}
\usepackage{array}
\usepackage{colortbl}

\usepackage{algorithm}
\usepackage{algpseudocode}
\usepackage{listings}

\usepackage{microtype}
\usepackage{xcolor}
\usepackage{soul}
\sethlcolor{yellow}

\usepackage{url}
\usepackage{doi}
\usepackage{natbib}

\usepackage{authblk}

\usepackage{orcidlink}

\usepackage{hyperref}
\hypersetup{
    colorlinks=true,
    linkcolor=black,
    filecolor=black,
    citecolor=blue,
    urlcolor=blue,
}
\newcommand{\undertitle}{}

\title{HELO Cryptography: A Lightweight Cryptographic System for Enhancing IoT Security in P2P Data Transmission}

\author[a]{Tahsin Ahmed\orcidlink{0009-0006-9820-1071}*\thanks{Corresponding author: Tahsin Ahmed\orcidlink{0009-0006-9820-1071}, Email: \href{mailto:tahsin.ahmed@g.bracu.ac.bd}{tahsin.ahmed@g.bracu.ac.bd}, URL: \href{https://www.tahsinahmed.com}{www.tahsinahmed.com}}}

\author[a]{Arjita Saha\orcidlink{0009-0004-1695-3890}}
\author[a]{Arian Nuhan\orcidlink{0009-0001-8034-2128}}
\author[a]{Nafim Ahmed Bin Mohammad Noor\orcidlink{0009-0000-7180-1639}}
\author[a]{Md Faisal Ahmed\orcidlink{0009-0004-4508-6570}}
\author[a]{Muhammad Iqbal Hossain\orcidlink{0000-0002-0915-9291}}

\affil[a]{Computer Science \& Engineering, BRAC University, Bangladesh}

\renewcommand{\shorttitle}{\textit{HELO Cryptography}}

\begin{document}

\maketitle

\begin{abstract}
The recent surge in security concerns for IoT devices highlights the increasing threat of cryptographic vulnerabilities. These weaknesses can lead to unauthorized access, data breaches, and manipulation of device functions, compromising the privacy and security of both the devices and their users. Given the limited computational power of IoT devices, especially when handling large amounts of data, encrypting and transmitting data over insecure networks poses significant challenges. This situation not only heightens security risks and prolongs runtime, but also degrades performance and consumes more resources. To address these issues, a novel cryptographic system named HELO (Hybrid Encryption Lightweight Optimization) is proposed. It is hybridized and gives solid security against cryptographic cyberattacks. However, the research objective is to enhance the security level of IoT devices without decreasing their performance. This system is ideal for resource-constrained gadgets due to its lightweight mechanism. Finally, it offers top-level cryptographic security for IoT gadgets by guaranteeing confidentiality, integrity, and availability while doing P2P data transmission.
\end{abstract}


\section{Introduction}\label{sec1}

\subsection{Overview}

The proliferation of IoT devices in different domains has promoted communication while at the same time raising challenges regarding data privacy and protection. IoT can be described as a network of physical objects embedded with sensors, software, and technologies for connectivity with the Internet to independently collect, share, and act on data, thus forming the basis of various automated systems. Examples of such devices include smartphones, wearable health monitors, smart home appliances, and industrial machinery. Thus, as IoT device integrations become more and more pervasive in day-to-day life, their security will be an important issue for preventing threats not only against things but also against the sensitive data that these devices will generate. As a result, cryptography is one of the very important factors in IoT systems for ensuring the confidentiality, integrity, and authenticity of data. However, these IoT devices are very vulnerable to cryptographic attacks due to their limited resources and difficulties in performing heavy computational tasks. So, lightweight cryptography therefore needs to be used as a countermeasure for these specific security challenges with minimum memory and computational overhead, as identified in \cite{bib20}.\\
\\
IoT nodes usually operate under highly constrained resources, such as low processing power and limited memory, which further makes them vulnerable to various security threats. Common vulnerabilities include exposure to personal information, data theft, compromised credentials, misconfigurations, broken access control, or even irreparable damage to reputation. Cryptographic failure can also cause heavy lawsuits. To address these challenges, robust cryptographic solutions tailored for IoT environments are essential, particularly in critical sectors such as healthcare, smart cities, industrial automation, and FinTech. Lightweight cryptography opens up prospective options in the areas of reduced computational power and energy consumption, lessened heat generation, and cost efficiency. It enables data protection against unauthorized access, integrity, and confidentiality for even resource-constrained IoT devices. Lightweight cryptography becomes an essential tool for balancing security with performance and power efficiency when securing IoT systems while attempting to maintain operational efficiency across various applications.

\subsection{Problem Statement}
Cryptographic algorithms typically demand high processing power to ensure security. While they perform well in resource-rich environments, they face significant challenges in resource-constrained settings like IoT. The key challenges include:
\begin{itemize}
\item \textbf{Decreased Security:} Traditional algorithms require extensive computational power but do not offer sufficient security in IoT environments, making systems vulnerable to unauthorized access.
\item \textbf{Energy Inefficiency:} Traditional cryptographic algorithms are energy-intensive, which is a significant drawback for IoT devices reliant on battery life. The high computational and memory demands lead to rapid battery depletion and overheating, which can be catastrophic in certain IoT apps.
\item \textbf{Compatibility Issues:} Many cryptographic models are incompatible with IoT devices due to their strict hardware and software requirements. They are not easily adaptable to the diverse needs of IoT systems, limiting their usability.
\item \textbf{Resource Strain:} IoT devices often have limited computational power and memory, making them unsuitable for traditional cryptographic algorithms that require significant resources. This strain on resources reduces the overall efficiency and functionality of IoT devices.
\item \textbf{Degraded Performance:} Traditional algorithms are complex, leading to slower processing and higher latency. This is especially problematic for IoT devices that require real-time processing.
\item \textbf{Cost Intensiveness:} The high hardware and resource requirements of traditional algorithms increase the cost of IoT device operation, making them impractical for widespread deployment.
\item \textbf{Key Management:} Managing cryptographic keys in large-scale IoT environments is a complex and critical task. If keys are compromised, the system becomes vulnerable to data breaches.
These challenges emphasize the need for lightweight, efficient cryptographic solutions tailored to the unique requirements of IoT devices.
\end{itemize}

\subsection{Research Objectives}
The research aim is to develop a highly secure cryptographic system for IoT users, which includes -

\begin{itemize}
    \item \textbf{Faster and Efficient Processing:} Ensuring optimal and faster performance of resource-constrained IoT devices. Also, ensuring the minimization of latency while focusing on efficient P2P data transmission for IoT devices.
    \item \textbf{Platform Independence:} Prioritizing hardware-independent cryptography to get rid of security vulnerabilities like cache-timing attacks, which is a major issue on AES. As a result, it eliminates expensive costs, which is an advantage in large-scale deployments.
    \item \textbf{Key Size Reduction:} Reducing the key size for encryption while maintaining security, addressing the limitations of algorithms like RSA.
    \item \textbf{Lightweight:} Building lightweight cryptography increases efficiency in resource-constrained devices that consume lower power, use minimal CPU and memory.
    \item \textbf{Enhaching Security:} Security is the major concern here. The primary objective is to enhance data security and privacy from multiple attacks without ignoring performance. 
\end{itemize}

\section{Thesis Organization}\label{sec2}
The rest of the thesis is organized as follows: Chapter 3 presents a literature review in which related works are discussed. Chapter 4 details the methodology and implementation of our proposed cryptographic algorithm. Chapter 5 is about the dataset that we used in our research. Chapter 6 shows the results and analysis of the experiment. Finally, Chapter 7 concludes the work, and Chapter 8 suggests a path for future research.

\section{Literature Review}\label{sec3}
\subsection{Related Work}
This paper discusses various lightweight cryptography techniques to secure IoT. It starts with a brief introduction to the layers of IoT architecture, which consists of application, communication, and physical layers. They are compared in terms of computational speed, storage, safety, and many other factors. The author has also mentioned the challenges of acquiring security in different layers of IoT \cite{8058307}.\\
\\
This study compares and contrasts three algorithms: AES, SPECK, and SIMON. The study uses the Cooja simulator on the Contiki operating system to test encryption methods in an IoT setting. The outcomes showed the pros and cons of each program in terms of how long it takes to run, how much power it uses, how much memory it takes up, and how fast it works \cite{10.4018/IJEHMC.2019100101}. \\
\\
The study divides lightweight cryptography into three groups: stream ciphers, block ciphers, and hash functions. It looks at how well cryptography algorithms work, their security, cost-effectiveness, and memory usage, as well as things like throughput, latency, memory usage, resistance to known attacks, expected time to break and die, and power consumption. Some important things to remember are how important lightweight cryptography is for protecting IoT devices with limited resources, how different algorithms work, how safe they are, how much they cost, and how the field needs more standardization. Also, energy efficiency is a crucial part of it \cite{9328432}.\\
\\
There is a full list of cryptographic algorithms that can be compared, such as Trivium, Elliptic Curve Cryptography (ECC), Tiny Encryption Algorithm (TEA), Proxy Re-Encryption, Attribute-Based Encryption (ABE), SM2 Cryptographic Schemes, 6LoWPAN, and IPSec. ECC and TEA are good choices for resource-constrained devices because of their small key sizes. However, Trivium is vulnerable to cubic attacks and necessitates a large hardware resource investment. ECC and TEA are frequently selected because of their effectiveness in resource-limited environments and authenticity \cite{9205259}, \cite{mhaibes2022simple}.\\
\\
Developing ECC for data key protection improves data integrity, and makes security stronger. The proposed system architecture uses ECC combined with an IoT platform for enhancing security. It won’t affect the size, performance, or energy consumption of devices. Also, the algorithm combines ECC with the Diffie-Hellman key exchange technique and AES for lightweight security. This algorithm uses key sizes of 256-bit and 512-bit and has implemented ECC curves that demonstrate how IoT Edge enhances security and increases processing capability for safe data transfer \cite{8795315}.\\
\\
Furthermore, ChaCha20, Poly1305, and MAC perform appropriately on various software platforms. The Authenticated Encryption with Associated Data (AEAD) scheme is created by combining the Poly1305 authenticator with the ChaCha20 stream cipher according to the RFC7539. The ChaCha20 stream cipher, Poly1305-ChaCha20 authenticator, and ChaCha20-Poly1305 AEAD scheme for ARM Cortex-M4 processors are all presented in the paper as compact, constant, and quick implementations. It handles 64-byte blocks, uses a secret key and nonce for encryption, has fast software performance, and ensures data encryption, integrity, and authenticity. Techniques for double round, quarter round, and Poly1305 multiplication optimization are discussed. ChaCha20 is optimized by switching the quarter-round order and reducing memory operations \cite{7927078}.\\
\\
This paper presents a survey regarding lightweight cryptographic solutions for IOT, including a comparison between hardware and the most recent software. In symmetric key cryptography, this survey includes security measures between different types of block ciphers such as AES, 3DES, Blowfish, Twofish, PRESENT, KATAN, TEA, Humming Bird, SIMON, and  RECTANGLE. Based on research, AES is considered a good security solution for restricted IoT devices. However, this paper also claims that AES for IOT can attract some hardware security attacks, such as side-channel attacks and correlation analysis attacks by using a false key and Wave Dynamic Differential Logic-based XOR gates. Asymmetric key cryptography includes RSA, DSA, and ECC. A comparative study shows that ECDH performs better than other algorithms in terms of power and area \cite{8666557}.\\
\\
This paper proposes a hybrid encryption algorithm, HAR (Hybrid AES Rail Fence), for secure data communication among IoT devices. This algorithm ensures confidentiality; the key generation process is carried out by the AES standard implementation, but does not show the way of key management approach. An attack called a biclique attack is faster than brute force by a factor of about four. Not only key-recovery attacks, but also side-channel attacks, padding-oracle attacks, and power-consumption analyses can be performed on AES. This shows that AES has many vulnerabilities. Ciphertexts produced by transposition ciphers are relatively easy to recognize because the frequency distribution of all ciphertext alphabet letters is the same as in plain messages written in the same language. This is why the Rail Fence Cipher can be broken quite easily by using brute force attacks due to the small number of possible keys. Lastly, compared with the AES algorithm, this hybrid approach resulted in a high avalanche effect, and provides more security in terms of complexity in finding data using diffusion techniques, but not in all methods to ensure strong security \cite{8907974}.\\
\\
Securing data could have been better in IoT. This is why the Blowfish algorithm has been proposed to enhance the security of IoT devices. It is modified where it depends on the rounds and substitution boxes, which speeds up the process of encryption and decryption. Its use of a 64-bit block size makes it vulnerable to birthday attacks, particularly in contexts like HTTPS. The GnuPG project recommends that Blowfish not be used to encrypt files larger than 4 GB due to its small block size \cite{9623559}.\\
\\
This paper proposes a hybrid algorithm combined with RSA and Diffie-Hellman. Diffie-Hellman acts as a secure key exchange protocol, and RSA accounts for securing the message. It shows analysis and comparisons between the traditional RSA algorithm and the proposed hybrid algorithm with some parameters like key size, message size, encryption, and decryption time. The encryption and decryption time of the hybrid algorithm is increased compared to the traditional RSA \cite{9132862}.

\section{Methodology}\label{sec4}
\subsection{Description of the Model}
This cryptographic model consists of seven mechanisms: asymmetric cryptography, symmetric cryptography, digital signature, data partitioning, tokenization, message authentication, and multithreading. These mechanisms enhance the security level even strongly than today’s adopted mechanisms for IoT devices. Now, let’s see how this cryptographic model works and securely handles data transmission.

\subsubsection{Chunk Size Initialization}
Data can sometimes be memory and CPU-intensive for a simple IoT device while doing cryptographic tasks. So, chunking data into several segments is a solution to this issue. This process allows partial decryption and data recovery without requiring access to the entire encrypted file. Also, it makes the model resilient to data corruption, where only the affected chunk will be lost rather than the entire dataset. This approach minimizes buffer overflow risks and reduces CPU overhead, resulting in faster encryption cycles and lower power consumption. Additionally, chunk-based processing improves transmission latency and fault tolerance, as corrupted or lost packets can be retransmitted independently without re-encrypting the entire dataset. However, a developer can give the size of each chunk \((C_s)\) according to their perspective. There is no block mode; instead, chunking data according to user preference gives freedom in development. So, this will increase efficiency, performance, and scalability in the constrained architecture of real-world IoT systems. Moreover, it becomes very beneficial for integrity checking and error handling (Algorithm \ref{algo1}).

\begin{algorithm}
\caption{Chunk Size Initialization}
\label{algo1}
\resizebox{1.0\textwidth}{!}{
\begin{minipage}{\textwidth}
\begin{enumerate}
    \item \textbf{Function} Chunk\_Size()
    \item \hspace*{1em} \textbf{Input:} \(C_s\) $\leftarrow$ "Enter chunk size"
    \item \hspace*{1em} \textbf{Return} \(C_s\) as integer
    \item \textbf{End Function}
\end{enumerate}
\end{minipage}
}
\end{algorithm}

\subsubsection{Key Generation}
\begin{figure}[htbp]
    \centering
    \includegraphics[width= 10.5cm, height= 11.5cm]{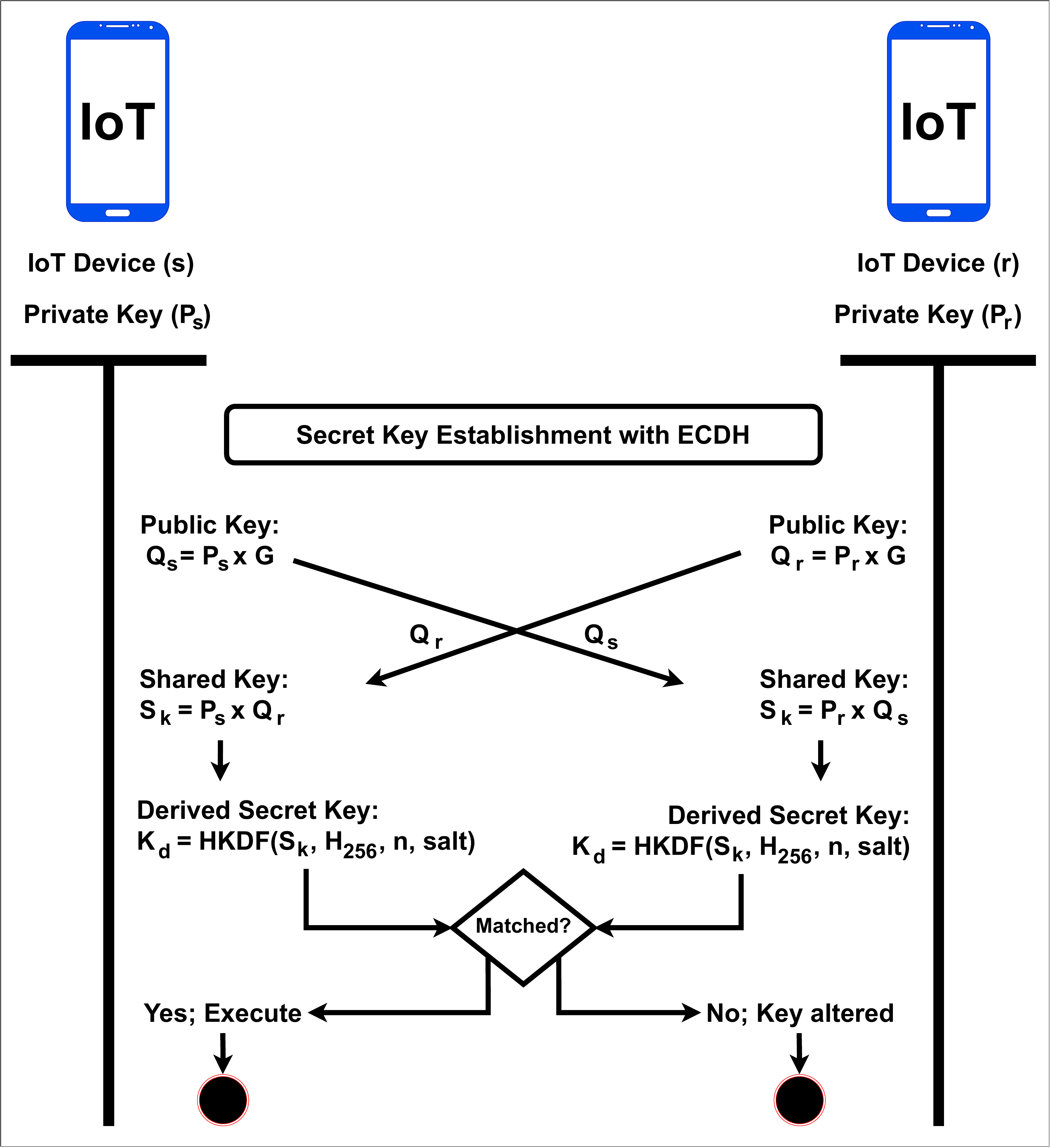}
    \caption{Workflow of  Key Exchange}
    \label{key_generation}
\end{figure}

The sender and the receiver generate a private and public key pair using the SECP25R1 elliptic curve. The formula of ECC is \(y^2=x^3+ax+b\) where SECP25R1 has a 256-bit point (x, y), the private key is a 256-bit scalar value \(a\), and it gives a public key point of \(a.G\). Furthermore, \(G\) is a base point of the curve, and it is predefined. It is chosen carefully so that the curve has a large cyclic subgroup with a high-order prime number. Here, the sender generates the public key \((Q_s)\) by multiplying its private key \((P_s)\) and \(G\). In addition, the receiver also generates the public key \((Q_r)\) by multiplying its private key \((P_r)\) and \(G\) (Algorithm \ref{algo2}).\\
\\
However, ECDH (Elliptic-Curve-Diffie-Hellman) is utilized in this model, allowing two parties to securely generate a shared secret key \((S_k)\) using their respective public and private keys. For verification on the sender's end, it takes three arguments: \(P_s\), \(Q_r\), and salt. Likewise, the receiver end takes three arguments: \(P_r\), \(Q_s\), and salt. This raw shared secret key is processed using the HMAC-based Key Derivation Function with SHA3-256 as the hash algorithm \((H_{256})\) and generates a derived key \((K_d)\). Also, the length \((n)\) for the key is 32 bytes or 256 bits. Now, both parties will exchange the hashed keys and check whether they match or not. The procedure will halt its execution if the keys are unmatched (Figure \ref{key_generation}) (Algorithm \ref{algo3}).

\begin{algorithm}
\caption{Generate Key Pair}
\label{algo2}
\resizebox{1.0\textwidth}{!}{
\begin{minipage}{\textwidth}
\begin{enumerate}
    \item \textbf{Function} Generate\_Keypair()
    \item \hspace*{1em} privateKey $\leftarrow$ generate\_private\_key(using SECP256R1)
    \item \hspace*{1em} publicKey $\leftarrow$ generate\_public\_key(privateKey)
    \item \hspace*{1em} \textbf{Return} (privateKey, publicKey)
    \item \textbf{End Function}
\end{enumerate}
\end{minipage}
}
\end{algorithm}

\begin{algorithm}
\caption{Key Exchange}
\label{algo3}
\resizebox{1.0\textwidth}{!}{
\begin{minipage}{\textwidth}
\begin{enumerate}
    \item \textbf{Function} Key\_Exchange(privateKey, publicKey, salt)
    \item \hspace*{1em} \(S_k\) $\leftarrow$ calculate\_shared\_secret(privateKey, publicKey)
    \item \hspace*{1em} \(K_d\) $\leftarrow$ derive\_key(\(S_k\), \(H_{256}\), salt, \(n\), info = ``handshake data'')
    \item \hspace*{1em} \textbf{Return} \(K_d\)
    \item \textbf{End Function}
\end{enumerate}
\end{minipage}
}
\end{algorithm}

\subsubsection{Tokenization}
Tokenization is applied when sensitive data needs to be replaced with unique identification symbols. Nonce \((T_n)\) is used in HELO as a token that will be generated once at a time, and the token will become invalid after that. As a result, its persistence becomes stateless and makes it unique per session. Here, the length \((n)\) of nonce is 16 bytes or 128 bits (Algorithm \ref{algo4}).

\begin{algorithm}
\caption{Tokenization}
\label{algo4}
\begin{enumerate}
    \item \textbf{Function} Generate\_Random\_Token()
    \item \hspace*{1em} \(T_n\) $\leftarrow$ generate\_random\_bytes(\(n\))
    \item \hspace*{1em} \textbf{Return} \(T_n\)
    \item \textbf{End Function}
\end{enumerate}
\end{algorithm}

\subsubsection{Message Authentication}
\begin{figure}[htbp]
    \centering
    \includegraphics[width=\columnwidth]{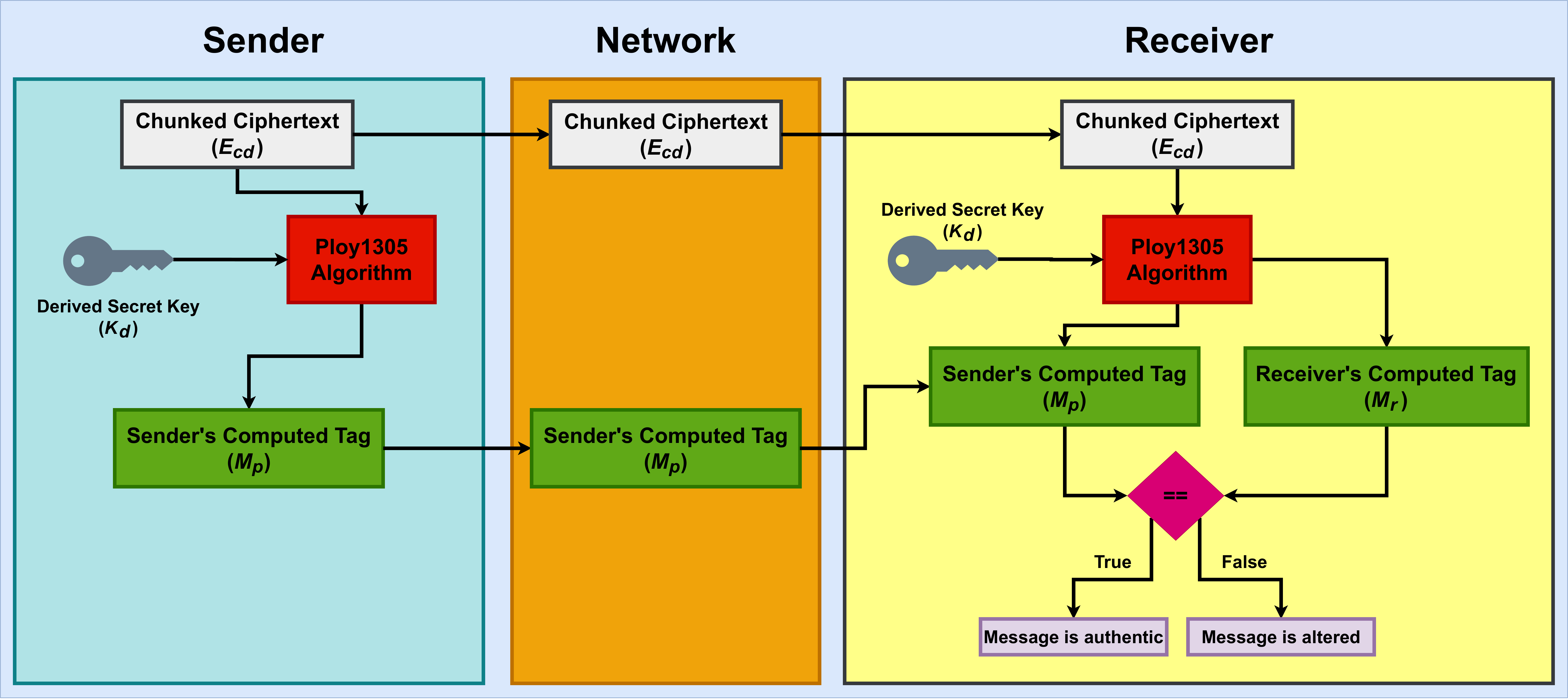}
    \caption{Workflow of Message Authentication}
    \label{Message Authentication Code}
\end{figure}

MAC (Message Authentication Code) is a security code that is employed to provide authentication of the origin of the data. Here, HELO uses Poly1305 algorithm \((P_{1305})\) as a MAC. The \(Generate\_Tag()\) function takes the sender’s derived shared secret key \((K_d)\), chunked ciphertext \((E_{cd})\) as inputs, and produces a 16-byte computed tag. The key size for the \(P_{1305}\) is 256 bits. In verification, the provided tag will be verified by the \(Verify\_Tag()\) function when the message reaches the receiver end. It will take three arguments: the receiver's derived shared secret key \((K_d)\), the chunked ciphertext \((E_{cd})\), and the sender's provided tag \((M_p)\). If the receiver's computed tag \((M_r)\) does not match with the sender’s provided tag \((M_p)\), it will generate an error instantly and exit the program (Figure \ref{Message Authentication Code}) (Algorithm \ref{algo5}).

\begin{algorithm}
\caption{Message Authentication}
\label{algo5}
\resizebox{0.95\textwidth}{!}{
\begin{minipage}{\textwidth}
\begin{enumerate}
    \item \textbf{Function} Generate\_Tag(\(K_d\), \(E_{cd}\))
    \item \hspace*{1em} \(M_p\) $\leftarrow$ compute\_mac(\(S_k\), \(E_{cd}\), \(P_{1305}\))
    \item \hspace*{1em} \textbf{Return} \(M_p\)
    \item \textbf{End Function}
    \item \textbf{Function} Verify\_Tag(\(K_d\), \(E_{cd}\), \(M_p\))
    \item \hspace*{1em} \(M_r\) $\leftarrow$ compute\_mac(\(K_d\), \(E_{cd}\), \(P_{1305}\))
    \item \hspace*{1em} \textbf{Return} \(M_r\) == \(M_p\)
    \item \textbf{End Function}
\end{enumerate}
\end{minipage}
}
\end{algorithm}

\subsubsection{Digital Signature}
\begin{figure}[htbp]
    \centering
    \includegraphics[width=\columnwidth, height=8cm]{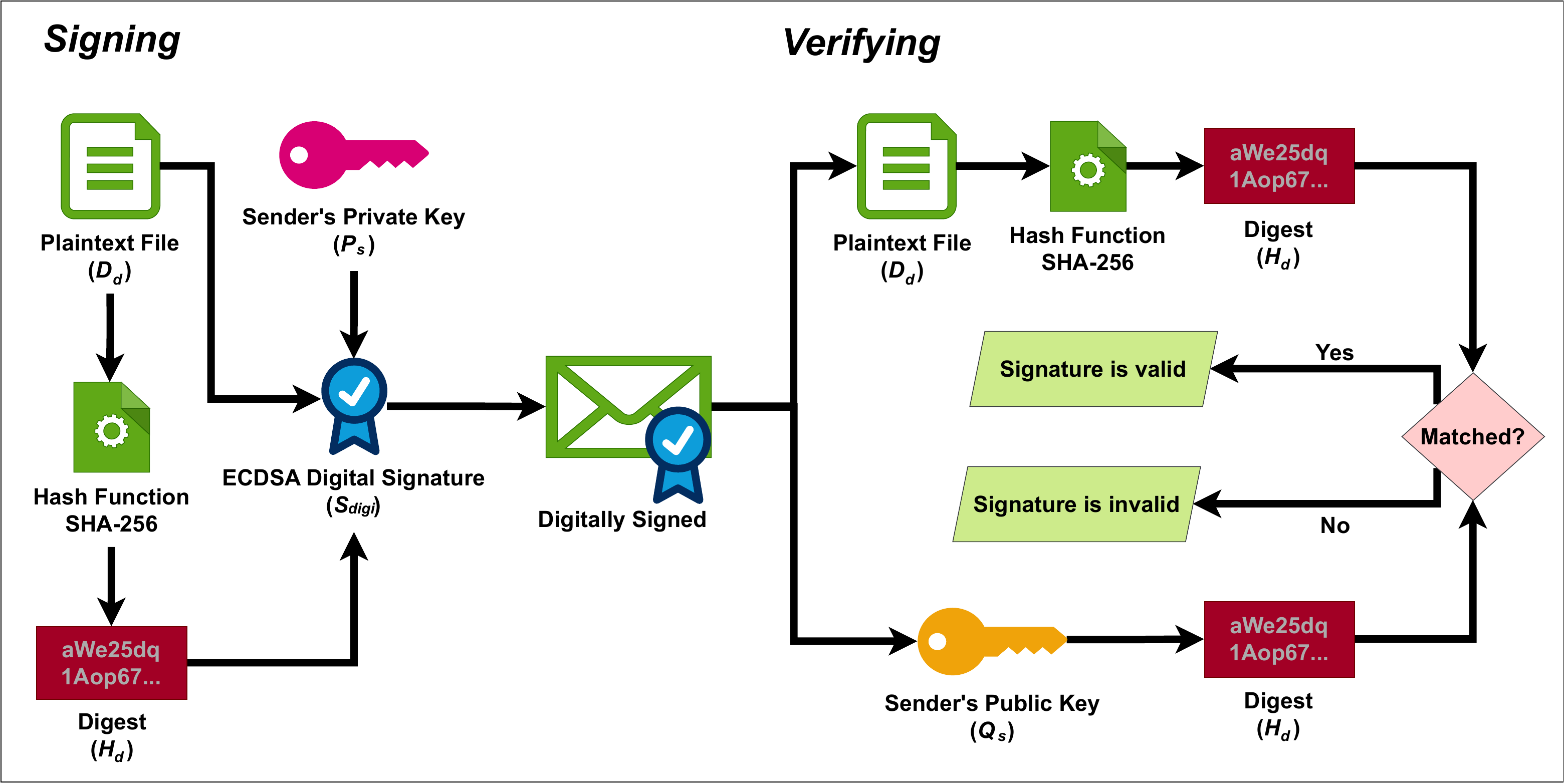}
    \caption{Workflow of Digital Signature}
    \label{Digital Signature}
\end{figure}

HELO uses ECDSA (Elliptic Curve Digital Signature Algorithm) to ensure the identity of an authentic sender. It digitally signs the plaintext before the encryption and sends it to its receiver through the network. Firstly, the \(Sign\_Data()\) function takes the sender’s private key \((P_s)\) and plaintext file \((D_d)\) as input. Now, a digest is generated by hashing the plaintext file using the SHA3-256 hash algorithm \((H_d)\). It then generates a signature \((S_{digi})\) by signing the data using the ECDSA algorithm by taking three parameters, which are the sender's public key \((Q_s\)), plaintext file, and generated digest \((H_d)\).\\
\\
In verification, the \(Verify\_Signature()\) function validates the integrity and authenticity of the data by verifying the signature against the provided public key \((Q_s)\) and plaintext file using the same ECDSA algorithm. In addition, a digest will be found when the receiver decrypts the digital signature using the sender's public key. Also, the receiver will again hash the plaintext file, generate a digest using the same SHA3-256 hash algorithm the sender used, and compare the result with the digest he received. It returns true if the verification process becomes successful. Otherwise, \(InvalidSignature\) will catch an exception and return false. It will then stop executing forward as it indicates possible tampering and forgery (Figure \ref{Digital Signature}) (Algorithm \ref{algo6}).

\begin{algorithm}
\caption{Digital Signature}
\label{algo6}
\resizebox{0.97\textwidth}{!}{
\begin{minipage}{\textwidth}
\begin{enumerate}
    \item \textbf{Function} Sign\_Data(\(P_s\), \(D_d\))
    \item \hspace*{1em} \(S_{digi}\) $\leftarrow$ Sign(\(P_s\), \(D_d\), \(H_d\))
    \item \hspace*{1em} \textbf{Return} \(S_{digi}\)
    \item \textbf{End Function}
    \item \textbf{Function} Verify\_Signature(\(Q_s\), \(D_d\), \(S_{digi}\))
    \item \hspace*{1em} \textbf{Try}
    \item \hspace*{2em} Validate\_Signature(\(S_{digi}\), \(Q_s\), \(D_d\), \(H_d\))
    \item \hspace*{2em} \textbf{Return} true
    \item \hspace*{1em} \textbf{Catch InvalidSignature}
    \item \hspace*{2em} \textbf{Print} ``Alert: Signature has been tampered.''
    \item \hspace*{2em} terminate\_execution()
    \item \hspace*{1em} \textbf{End Try}
    \item \textbf{End Function}
\end{enumerate}
\end{minipage}
}
\end{algorithm}

\subsubsection{Encryption}
\begin{figure}[H]
    \centering
    \includegraphics[width=\columnwidth]{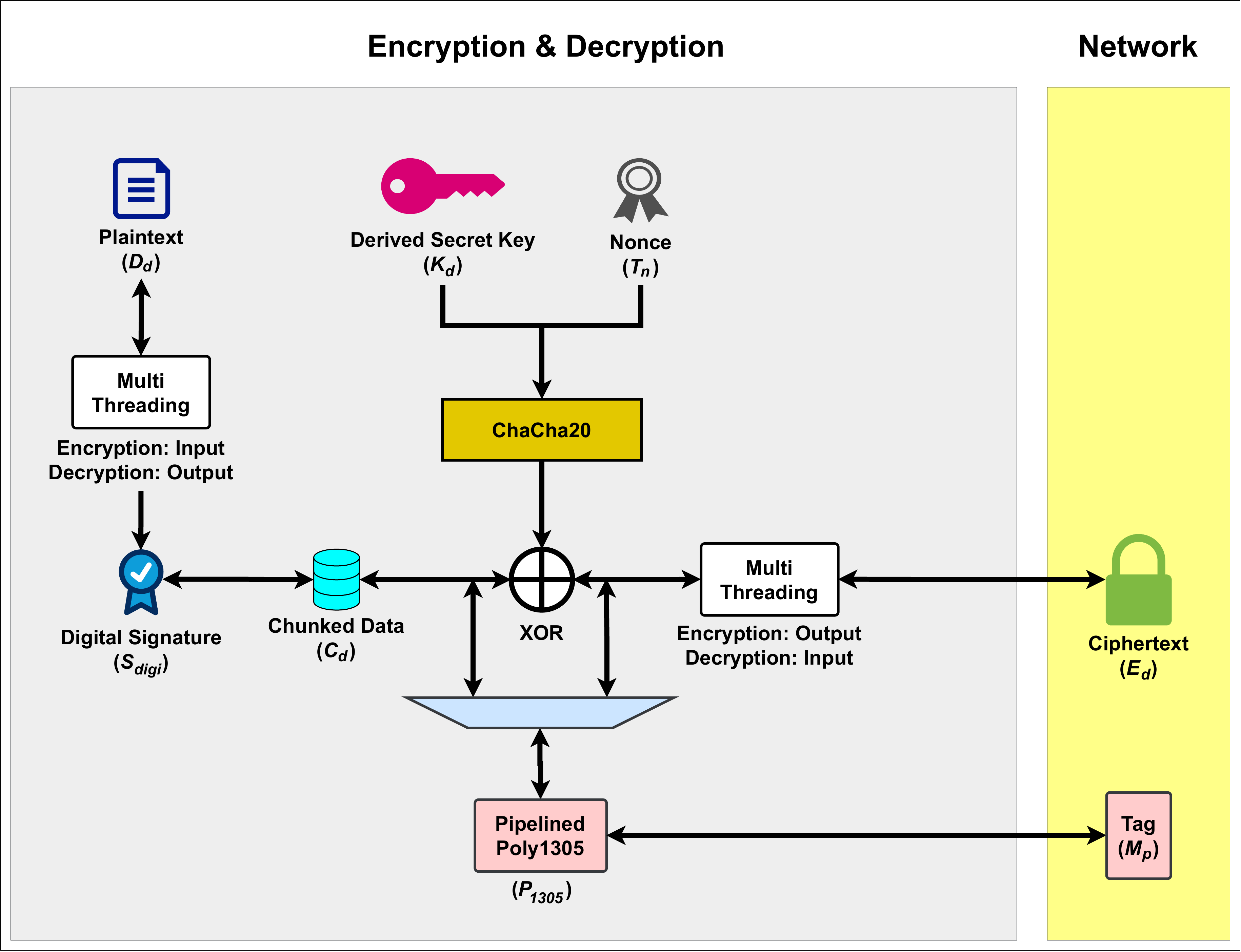}
    \caption{Workflow of Encryption and Decryption}
    \label{Encryption and Decryption}
\end{figure}

Encryption is responsible for transforming information into an unreadable ciphertext. Firstly, it starts with determining the chunk size using \(Chunk\_Size()\) function (Algorithm \ref{algo1}), although this variable is not directly used in the encryption, but rather in a chunk-based operation. The plaintext file \((D_d)\) is converted into a binary mode before the XOR operation. Secondly, ECDSA \((S_{digi})\) is applied in the plaintext file using the sender's private key \((P_s)\), ensuring the authenticity and integrity of the plaintext data. For later verification, it is stored in a separate file with a \(.sig\) extension. Afterward, the binary mode of the plaintext file is chunked \((C_d)\). Furthermore, a nonce \((T_n)\) is generated using \(Generate\_Random\_Token()\) function (Algorithm \ref{algo4}) to initialize the ChaCha20 cipher algorithm \((C_i\)).\\
\\
Here, the cipher is created with a nonce and the sender's derived secret key \((K_d)\). The chunked plaintext is then encrypted using the cipher's encryptor \((Enc_c\)) and generates ciphertext. This ciphertext \((E_{cd})\) is now finally finalized and stored. Furthermore, \(Generate\_Tag()\) function (Algorithm \ref{algo5}) generates a MAC using Poly1305 \((P_{1305})\), which is pipelined with the XOR operation of the ChaCha20 algorithm \((Cha_{20}\)) to ensure the integrity of each of the chunked ciphertexts. This encrypted data, along with the sender's computed tag \((M_p)\), forms the payload and is transmitted through the network. The encrypted file contains nonce, tag, and ciphertext. However, multithreading plays a vital role in handling the transmission of data more efficiently during encryption (Figure \ref{Encryption and Decryption}) (Algorithm \ref{algo7}).\\
\\
ChaCha20 is a stream cipher that encrypts data byte-by-byte without adding any padding characters, which means the size of the chunked ciphertext (\(M)\) is exactly as same as the size of the chunked plaintext. So, the total size of the encrypted file \((E_{cd})\) is calculated as follow: \\

\noindent
\fontsize{12pt}{15pt}\selectfont
\normalsize
\(
\begin{aligned}
\text{Total size} &= \text{size of \(T_n\)} + \text{size of \(E_{cd}\)} + \text{size of \(M_p\)} \\
                  &= 16 + M + 16 \\
                  &= M + 32 \text{ bytes}
\end{aligned}
\)
\normalsize

\begin{algorithm}[htbp]
\caption{Encrypt Chunked File}
\label{algo7}
\scalebox{1}{
\begin{minipage}{\textwidth}
\begin{enumerate}
    \item \textbf{Function} Encrypt\_Chunked\_File(\(D_d\), \(E_d\), \(K_d\), \(P_s\))
    \item \hspace*{1em} \(C_d\) $\leftarrow$ Chunk\_Size()
    \item \hspace*{1em} \textbf{Open} \(D_d\) for \textbf{Reading}
    \item \hspace*{2em} \(S_{digi}\) $\leftarrow$ Sign\_Data(\(P_s\), \(D_d\))
    \item \hspace*{2em} \(T_n\) $\leftarrow$ Generate\_Random\_Token()
    \item \hspace*{2em} \(C_i\) $\leftarrow$ initialize\_cipher(\(Cha_{20}\), \(K_d\), \(T_n\))
    \item \hspace*{2em} \(Enc_c\) $\leftarrow$ create\_encryptor(\(C_i\))
    \item \hspace*{2em} \textbf{While} data\_chunk $\leftarrow$ read\_chunk(\(D_d\), \(C_d\)) \textbf{do}
    \item \hspace*{3em} \(E_{cd}\) $\leftarrow$ encryptor.encrypt(data\_chunk)
    \item \hspace*{3em} \(F_e\) $\leftarrow$ finalize\_encryption(\(E_d\), \(Enc_c\))
    \item \hspace*{3em} \(M_p\) $\leftarrow$ Generate\_Tag(\(K_d\), \(F_e\))
    \item \hspace*{2em} \textbf{End While}
    \item \hspace*{2em} \textbf{Open} \(E_d\) for \textbf{Writing}
    \item \hspace*{3em} \textbf{Write} \(E_{cd}\) to \(E_d\)
    \item \hspace*{3em} \textbf{Write} \(M_p\) to \(E_d\)
    \item \hspace*{3em} \textbf{Write} \(T_n\) to \(E_d\)
    \item \hspace*{2em} \textbf{Close} \(E_d\)
    \item \hspace*{1em} \textbf{Close} \(D_d\)
    \item \hspace*{1em} \textbf{Open} \(E_d\) for \textbf{Writing}
    \item \hspace*{2em} \textbf{Write} \(S_{digi}\) to (\(E_d\) + .sig)
    \item \hspace*{1em} \textbf{Close} \(E_d\)
    \item \textbf{End Function}
    \end{enumerate}
\end{minipage}
}
\end{algorithm}

\subsubsection{Threading}
\begin{figure}[htbp]
    \begin{flushleft}
    \includegraphics[width=18cm, height=19cm]{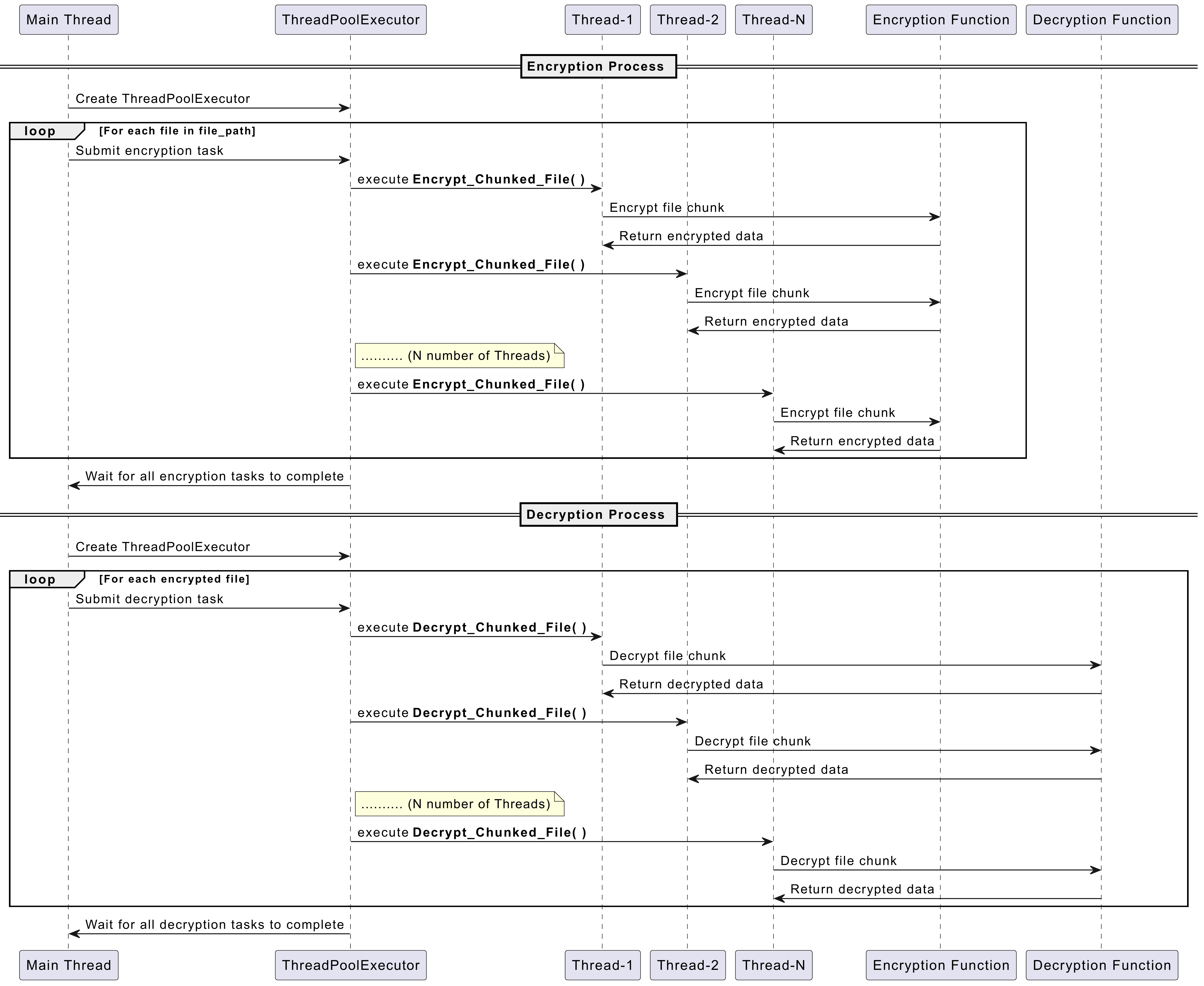}
    \end{flushleft}
    \caption{Workflow of Multithreading}
    \label{Multithreading}
\end{figure}

This model utilizes multithreading for improved resource efficiency, as threads are lighter than processes and maximize CPU usage. By running tasks concurrently, threading ensures faster operations for battery-operated IoT devices. Here, multithreading enhances encryption and decryption efficiently, where a single process is divided into multiple threads. Firstly, \(Threaded\_File\_Encryption\) function takes four parameters: a path of the source file \((S_{path})\), a path where the encrypted file will be stored \((E_{path})\), a derived secret key, and the sender's private key. Secondly, a thread pool executor \((T_{pool})\) has to be initialized. Thirdly, it takes an empty list \((L_o)\) for appending the objects to keep track of all running tasks for encryption. Afterward, it provides encryption for all files in a specified folder by iterating through them and sending each file to a \(T_{pool}\) for concurrent processing. It creates the total path for the plaintext file and its corresponding encrypted file, adding a \(.enc\) extension to point out the encryption. Each file is handled at the point utilizing \(Encrypt\_Chunked\_File()\) function (Algorithm \ref{algo7}). It performs encryption using \(D_d\), \(E_d\), sender's derived secret key \((K_d)\) and his private key \((P_s)\). Finally, \(T_{pool}.wait\_for\_all\_threads()\) function guarantees that all encryption tasks are complete before the function's exit (Algorithm \ref{algo8}).\\
\\
Likewise, the \(Threaded\_File\_Decryption()\) function is responsible for decrypting all files from the encrypted folder concurrently. Firstly, it takes four arguments: the path of the encrypted file \((E_{path})\), a path where the decrypted file will be stored \((D_{path})\), receiver's derived secret key \((K_d)\), and sender's public key \((P_s)\). It also checks for files ending with the \(.enc\) extension, which helps the program to remove it and reconstruct the decrypted file. Secondly, the thread pool executor becomes initialized for multithreading, and the decrypted file is then submitted to it. Afterward, \(Threaded\_File\_Decryption()\) function takes an empty list \((L_o)\) for appending the objects in terms of keeping track of all running tasks for decryption, which is handled by the \(Decrypt\_Chunked\_file\) function (Algorithm \ref{algo10}). Finally, again the function \(T_{pool}.wait\_for\_all\_threads()\) guarantees that all decryption tasks are complete before the exit of the function (Figure \ref{Multithreading}) (Algorithm \ref{algo9}).

\begin{algorithm}
\caption{Threading for Encryption}
\label{algo8}
\scalebox{0.98}{
\begin{minipage}{\textwidth}
\begin{enumerate}
    \item \textbf{Function} 
    Threaded\_File\_Encryption(\(S_{path}\), \(E_{path}\), \(K_d\), \(P_s\))
    \item \hspace*{1em} Initialize \(T_{pool}\) for multithreading
    \item \hspace*{1em} \(L_o\) $\leftarrow$ [ ]
    \item \hspace*{1em} \textbf{For} file\_name \textbf{in} file\_path \textbf{do}
    \item \hspace*{2em} \(D_d\) $\leftarrow$ \textbf{concat} \(S_{path}\) \textbf{and} file\_name
    \item \hspace*{2em} \(E_d\) $\leftarrow$ \textbf{concat} \(E_{path}\) \textbf{and} file\_name + .enc
    \item \hspace*{2em} \textbf{Append}(\(L_o\), \(T_{pool}\).create\_thread(\textbf{call} Encrypt\_Chunked\_File(\(D_d\), \(E_d\), \(K_d\), \(P_s\))))
    \item \hspace*{1em} \textbf{End For}
    \item \hspace*{1em} \(T_{pool}\).wait\_for\_all\_threads()
    \item \textbf{End Function}
\end{enumerate}
\end{minipage}
}
\end{algorithm}

\begin{algorithm}
\caption{Threading for Decryption}
\label{algo9}
\scalebox{0.98}{
\begin{minipage}{\textwidth}
\begin{enumerate}
    \item \textbf{Function} Threaded\_File\_Decryption(\(E_{path}\), \(D_{path}\), \(K_d\), \(Q_s\))
    \item \hspace*{1em} Initialize \(T_{pool}\) for multithreading
    \item \hspace*{1em} \(L_o\) $\leftarrow$ [ ]
    \item \hspace*{1em} \textbf{For} file\_name \textbf{in} \(E_{path}\) \textbf{do}
    \item \hspace*{2em} \(E_d\) $\leftarrow$ \textbf{concat} \(E_{path}\) \textbf{and} file\_name
    \item \hspace*{2em} \(D_d\) $\leftarrow$ \textbf{concat} \(D_{path}\) \textbf{and} remove .enc from file\_name
    \item \hspace*{2em} \textbf{Append}(\(L_o\), \(T_{pool}\).create\_thread(\textbf{call} Decrypt\_Chunked\_File(\(E_d\), \(D_d\), \(K_d\), \(Q_s\))))
    \item \hspace*{1em} \textbf{End For}
    \item \hspace*{1em} \(T_{pool}\).wait\_for\_all\_threads()
    \item \textbf{End Function}
\end{enumerate}
\end{minipage}
}
\end{algorithm}

\begin{algorithm}
\caption{Decrypt Chunked File}
\label{algo10}
\scalebox{0.98}{
\begin{minipage}{\textwidth}
\begin{enumerate}
    \item \textbf{Function} Decrypt\_Chunked\_File(\(E_d\), \(D_d\), \(K_d\), \(Q_s\))
    \item \hspace*{1em} \textbf{Open} \(E_d\) for \textbf{Reading}
    \item \hspace*{2em} \(T_n\) $\leftarrow$ extract\_first\_bytes(\(E_d\), 16)
    \item \hspace*{2em} \(M_r\) $\leftarrow$ extract\_last\_bytes(\(E_d\), 16)
    \item \hspace*{2em} \(E_{cd}\) $\leftarrow$ extract\_remaining\_bytes(excluding = \(T_n\) and \(M_r\))
    \item \hspace*{2em} \textbf{If} Verify\_Tag(\(K_d\), \(E_{cd}\), \(M_r\)) \textbf{then}
    \item \hspace*{3em} cipher $\leftarrow$ initialize\_cipher(\(C_i\), \(K_d\), \(T_n\))
    \item \hspace*{3em} decryptor $\leftarrow$ create\_decryptor(\(E_{cd}\))
    \item \hspace*{3em} \textbf{Open} \(D_d\) for \textbf{Writing}
    \item \hspace*{4em} \textbf{While} \(C_d\) $\leftarrow$ decryptor.decrypt(cipher)
    \item \hspace*{5em} \textbf{Write} \(C_d\) to \(D_d\)
    \item \hspace*{5em} finalize\_and\_write(\(D_d\), decryptor)
    \item \hspace*{5em} Verify\_Signature(\(Q_s\), \(D_d\), \(E_d\) + .sig)
    \item \hspace*{4em} \textbf{End While}
    \item \hspace*{3em} \textbf{Close} \(D_d\)
    \item \hspace*{2em} \textbf{Else}
    \item \hspace*{3em} \textbf{Print} ``Alert: Tags are mismatch.''
    \item \hspace*{3em}  terminate\_execution()
    \item \hspace*{2em} \textbf{End If}
    \item \hspace*{1em} \textbf{Close} \(E_d\)
    \item \textbf{End Function}
\end{enumerate}
\end{minipage}
}
\end{algorithm}

\subsubsection{Decryption}
Decryption is the reverse process of encryption, which converts the encrypted data back to its original form and is also human-readable.\\
\\
Likewise, in our proposed cryptographic system, the decryption phase starts by opening the encrypted file \((E_d)\) in binary mode and reading the first 16 bytes as nonce \((T_n)\). The last 16 bytes of the data are extracted as the sender's provided tag \((M_p)\). The rest of the data consists of chunked ciphertext \((E_{cd})\). After successfully authenticating the ciphertext with the receiver's computed tag \((M_r)\), the ChaCha20 algorithm decrypts the ciphertext file using the nonce, and the receiver’s derived secret key \((K_d)\) which must be matched with the sender’s derived secret key \((K_d)\). Finally, decryption proceeds using the ChaCha20 algorithm \((C_i\)) unless the nonce expires, allowing the receiver to successfully receive the original file. After decryption, the \(Verify\_Signature()\) function verifies the sender's genuineness by checking his digital signature using the ECDSA algorithm. The signature is stored in an isolated record with a \(.sig\) extension \((E_d+.sig)\). It is read and validated against plaintext utilizing the sender's public key \((Q_s)\). If both tag and signature verification pass, decrypted plaintext \((C_d)\)\ is been written to a new file \((D_d)\), ensuring that only an authorized user can access the original content (Figure \ref{Encryption and Decryption}) (Algorithm \ref{algo10}).

\subsection{Implementation}

\begin{figure}[htbp]
    \includegraphics[width=18cm, height=11.5cm]{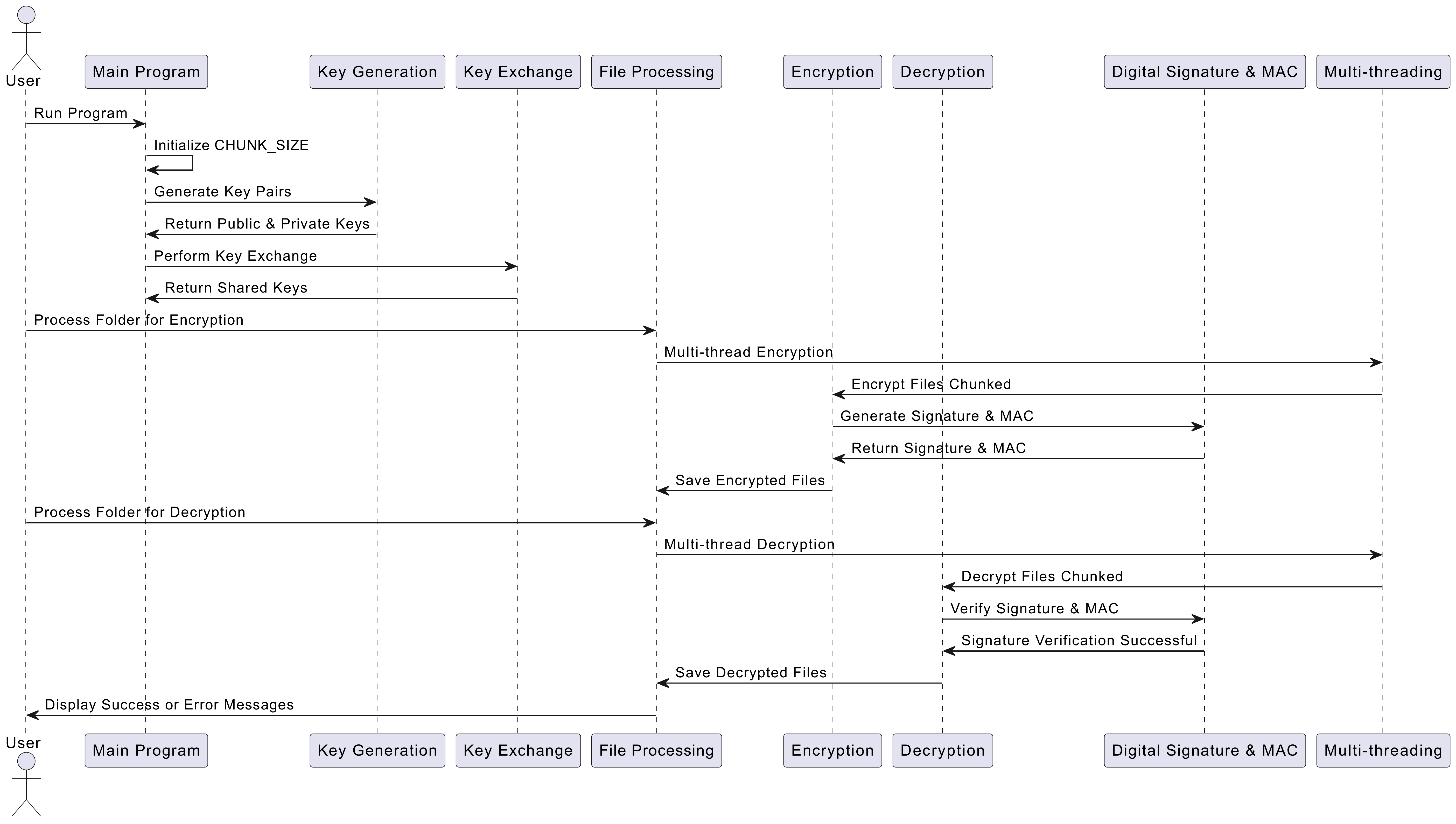}
    \caption{UML Sequence Diagram of Implementing HELO Cryptography}
    \label{UML Sequence Diagram of HELO Implementation}
\end{figure}

\subsubsection{Experimental Setup}
All experimental evaluations of the proposed lightweight cryptographic scheme were performed on a personal desktop computer instead of an actual IoT device. The system configuration includes an Intel(R) Core(TM) i5-9600K processor, which has 6 cores and threads, 9 MB cache, and a base frequency of 3.70 GHz. Here, the base frequency is the operating point where the TDP (Thermal Design Power) is defined. Moreover, the TDP value is 95 Watts, referring to the maximum amount of heat a CPU is expected to generate under full load. Also, 16 GB of DDR4-2666 RAM is used, and the experiment is performed on the Microsoft Windows 11 Pro operating system. Additionally, several Python libraries are used to analyze performance, the avalanche effect, and energy consumption. Also, the Flask framework is employed for the web interface.

\subsubsection{Limitations}
The mentioned configuration introduces certain limitations. Specifically, the absence of an IoT device, such as a resource-constrained Raspberry Pi Zero, ESP32, or Arduino Nano 33, restricts the generalization of the results to embedded environments. Furthermore, the estimated results were derived from energy consumption rather than hardware-level power measurements. This experiment was conducted on a single machine. Consequently, factors such as scalability, distributed processing, and computational overhead analysis became increasingly difficult in large-scale IoT deployments. As a result, we had to analyze the information hypothetically and quantitatively.

\subsubsection{Software Design}
A UML diagram is created for the HELO cryptography software application. It is designed to run on a web server and be accessed by users through a web browser. Additionally, it can be described as a web application (Figure \ref{UML Sequence Diagram of HELO Implementation}). Furthermore, the system is designed to facilitate the encryption and decryption of the files we have created. Additionally, the project's source code has been uploaded to GitHub and is open-source. As a result, this will create opportunities for developers and researchers to work on it.\\
\\
\textbf{Source Code:} \href{https://github.com/tahsinnahmed/HELO-Cryptography.git}{https://github.com/tahsinnahmed/HELO-Cryptography.git} \\
\textbf{Web Application:} \href{https://helocrypto.pythonanywhere.com}{https://helocrypto.pythonanywhere.com}

\section{Dataset Description}\label{sec5}
In this section, the experiments were conducted using datasets obtained from Kaggle, which were chosen because they capture diverse data patterns that closely resemble those found in real-world IoT environments. Although the evaluation was performed in a simulated desktop environment, the dataset effectively reflects an IoT-specific network of instructions and resource constraints. The incorporation of HELO Cryptography in the experimental framework further aligns with the computational and energy-efficient characteristics expected in practical IoT deployments. This is done to ensure that the results remain representative and applicable. As a result, the findings provide a valid indication of how the proposed cryptographic system approach would perform in real IoT systems under lightweight cryptographic conditions.

\begin{table*}[htbp]
\caption{ Dataset Description}
\label{tab1}
\resizebox{\textwidth}{!}{ 
\begin{tabular}{@{}lllll@{}}
\toprule
Dataset Type  & File Name & Size & Source & Key Features\\
\midrule
TXT    & 1k Random 3D Shapes  & 15 Bytes  & \cite{dataset_1} & Single line, seeding value  \\
     & TACO Dataset YOLO Format & 118 Bytes & \cite{dataset_2} & README file  \\
   & TF 2.16 Requirements & 171 Bytes & \cite{dataset_3} & Python package list \\
   & English to French Translations & 1 Kilobyte & \cite{dataset_4} & Terms conditions\\
   & Thai Sentiment Analysis Toolkit & 10 Kilobytes & \cite{dataset_6} & Thai Text Sentiment Analysis\\
    & Hip-Hop Encounters Data Science & 22 Kilobytes & \cite{dataset_5} & Song lyrics\\
\midrule
CSV & Netflix OTT Revenue and Subscribers & 1 Megabyte & \cite{dataset_7} & Banking transactions\\
 & Credit Score Classification Dataset & 6 Megabytes & \cite{dataset_8} & Financial records\\
 & FIFA 22 Complete Player Dataset & 10 Megabytes & \cite{dataset_9} & Football player profiles\\
 & Midjourney 2022 - 250k & 67 Megabytes & \cite{dataset_10} & Image generation metadata\\
 & World's Top 200,000 Scientists & 84 Megabytes & \cite{dataset_11} & Scientific contributions\\
 & Football Data from Transfermarkt & 124 Megabytes & \cite{dataset_12} & Football statistics\\
 \midrule
Images & Apples or Tomatoes Image Classification & 368 Kilobytes & \cite{dataset_13} & Labeled apple images\\
  & Soil Types & 600 Kilobytes & \cite{dataset_14} & Peat soil images\\
  & Cat Dog Images & 740 Kilobytes & \cite{dataset_15} & Cat and dog images\\
  & Pokemon Image Dataset & 5 Megabytes & \cite{dataset_16} & Pokémon images\\
    & Pistachio Image Dataset & 20 Megabytes & \cite{dataset_17} & 
    Pistachio images \\
  & Weather Image Recognition & 40 Megabytes & \cite{dataset_18} & Weather images \\
\bottomrule
\end{tabular}
}
\end{table*}

The collection of data contains three types, which are TXT, CSV, and image files (JPG, PNG). These datasets are primarily used for testing encryption and decryption algorithms in multiple formats. The structure of the data is well organized to keep consistency and fairness in check for comparisons. Below is the dataset information that was used (Table \ref{tab1}).

\section{Results and Analysis}\label{sec6}

\subsection{Attack Type}
When it comes to transmission, people have significant risks from a variety of cyber threats. The purpose is to identify the most common attack types that occur during data exchange and to provide a full picture of how it works and what the consequences are. These attacks show the importance of cybersecurity for the safety and privacy of transmission.

\begin{itemize}

\item \textbf{MITM Attack:} An attacker can listen in on or change data during the attack by standing between two people talking to each other \cite{5563900}.
\item \textbf{Spoofing:} Pretending to be a real person, group, or source to trick people or get unauthorized access \cite{Babu2010}.
\item \textbf{Supply Chain Attack:} To get into target systems or networks, hackers use flaws in third-party providers or services \cite{769ac4f6}. 
\item \textbf{Replay Attack:} To get unauthorized access or pretend to be a real user, hackers record and playback valid data transfers \cite{ALBINALI2025110996}.
\item \textbf{Brute Force Attack:} Try-and-fail way for guessing passwords or encryption keys to get in without permission \cite{app13105979}.
\item \textbf{Dictionary Attack:} Using a list of common words or sentences as passwords or to get into a system or account without permission \cite{app13105979}.
\item \textbf{Rainbow Attack:} Hashing methods are used to encrypt passwords and make hash tables that can be used to break them \cite{10.1007/11767480_13}.
\item \textbf{Eavesdropping:} Illegal communication interceptions carried out to obtain data or information \cite{s21248207}.
\item \textbf{Tampering:} Data, systems, or messages that have been changed without permission \cite{9117939}.
\item \textbf{Forgery:} Creating bogus or forged identities, digital signatures, or documents \cite{shim2019universal}. 
\item \textbf{Reverse Engineering:} Looking closely at an object or system to learn about its parts, how it works, or how it was designed \cite{8488542}.
\item \textbf{Birthday Attack:} Cryptographic hash collisions can be used to make two different inputs that have the same hash result \cite{chavan2023investigating}.
\item \textbf{Preimage Attack:} A cryptographic attack in which the attacker looks for input that
hashes to a particular value. Preimage attacks are classified into two types: first, which involves finding any input for a given hash, and second, which involves finding an alternative input that matches the hash of a given input \cite{aoki2009preimage}.
\item \textbf{Cache-Timing Attack:} A kind of side-channel assault in which an attacker takes advantage of differences in the amount of time it takes to access data stored in a computer’s cache. Based on how data is accessed and processed, an attacker can determine sensitive information, including cryptographic keys, by analyzing these temporal disparities \cite{weiss2012cache}.
\end{itemize}

\subsection{Attack Analysis}
To figure out which security methods can help stop different kinds of cyberattacks, it’s important to look at the pros and cons of each method in the context of the attacks. Here’s how our model cryptography approaches might mitigate various attacks:

\subsubsection{Nonce}
The term ‘nonce’ stands for ‘number used once’, and generates a random number using a Cryptographically Secure Pseudorandom Number Generator (CSPRNG) for each time or transaction that is difficult to detect, predict, and utilize in a short time. Its randomness, uniqueness, and limited validity enhance security, which can prevent replay attacks \cite{pandiya2020mitigating}.

\subsubsection{SHA3-256}
SHA3-256 uses a sponge structure based on the Keccak algorithm for establishing a 1600-bit state; it logs input blocks using XOR and Keccak-f permutations. The 256-bit hash value is extracted, making the architecture very resistant to collision attacks and difficult for attackers to break, which are also computationally costly and difficult to detect with present technology. This gives SHA3-256 protection against preimage attacks by requiring a distinct input with the same hash and complexity of \(2^{256}\). If an attacker knows the original message hash, they can calculate the extended message hash. Extending length Security is provided via SHA3-256. This method protects against differential cryptanalysis by reflecting changes in the hash output and being highly sophisticated. Also, birthday attack utilizes the probability theory birthday paradox to locate 2 messages with the same hash value faster than brute force where \(2^{256}\) complications make birthday attacks more difficult \cite{luo2016differential}.

\subsubsection{Elliptic Curve Diffie-Hellman (ECDH) and Elliptic Curve Digital Signature Algorithm(ECDSA)}

ECDH uses elliptic curves to make key sharing safer and allow two parties to agree via a protected channel where each party produces public and private keys and exchanges the public key. Adding those keys creates a shared secret key for secure communication using symmetric encryption. It protects communication by discussing secret keys without giving personal information. Furthermore, ECDSA generates and verifies digital signatures. Create a private-public key pair, sign a message with the private key, and then confirm with the public key, ensuring message integrity, non-repudiation, and authentication \cite{adarbah2023security}.\\
\\
ECDH and ECDSA combo handles many attacks, with ECDH for verification and ECDSA for key sharing for more secure and trustworthy communication. ECDSA (Elliptic Curve Digital Signature Algorithm) uses digital signatures to prevent spoofing, ensuring that only the holder of the private key can sign a message, which is verified by the corresponding public key. By confirming the legitimacy of the signatures, ECDSA may determine whether a key has been compromised or not. Without the private key, it is nearly impossible to forge signatures. It also reduces the risk of MITM (Man-in-the-Middle) attacks by using secure key exchange protocols like ECDH, preventing tampering, eavesdropping, and cryptanalytic attacks by utilizing forward secrecy. To ensure continued security, a compromised private key can be exchanged with a fresh key pair. ECDSA ensures message integrity and authenticity, making brute-force and supply-chain attacks impractical due to the complexity of elliptic curve cryptography \cite{farooq2019ecdsa}.

\subsubsection{ChaCha20 with Poly1305}
ChaCha20 is a stream of pseudo-random bits that is XORed with the plaintext to create the ciphertext. A shared secret key, counter, and nonce provide a secure pseudorandom key stream for data privacy and integrity. The protected message, data, nonce, and secret key are turned into a MAC, and an authentication tag using POLY1305 MAC certifies integrity and validity. If the computed and received tags match, the message is not altered. Now, using ChaCha20 and Poly1305 together ensures strong encryption and authentication, preventing hacking and interception of communication paths. Also, cryptography protects sensitive data from unauthorized access. It is difficult to reverse-engineer this mixture. It generates unique verification tags for each message, even with identical plaintexts. This prevents known attacks utilizing plaintext. Hackers cannot access encryption keys or sensitive communication data. Moreover, ChaCha20 with Poly1305 guards against specific plaintext attacks by utilizing Poly1305 for message authentication, guaranteeing that attackers cannot change or construct new ciphertexts without the secret key, even if they select plaintexts and access their ciphertexts. Also, as ChaCha20 is built to be resistant to timing attacks, it can defeat cache-timing attacks because of its data-independent and constant-time procedures; the encryption process takes the same amount of time regardless of input data. In addition, Poly1305’s safe authentication and ChaCha20’s strong encryption and security stem from its ability to withstand cryptographic attacks, such as key recovery attacks \cite{7927078}, \cite{9509883}.

\subsection{Performance Analysis}
Performance analysis is structured into three key components: CPU Processing Time, RAM Usage, and Runtime Analysis, across three file types. These are TXT, CSV, and image files. This methodology ensures a consistent evaluation of algorithms and validates their capabilities across various file types, providing confidence in their performance. Also, CPU processing time refers to the time the CPU takes to execute instructions from a program. The total time from the start to the completion of program execution is known as the runtime. Moreover, the amount of Random Access Memory consumed by a program during execution is referred to as RAM usage.\\
\\
Here, analyzing the parameters is crucial for cryptographic algorithms, as they impact efficiency and performance. Optimized CPU time leads to faster operations essential for real-time processing, while minimized runtime ensures scalability under high operational volumes. Also, Efficient RAM usage is vital in resource-constrained environments like IoT devices. Furthermore, consistent operation times and secure memory management can help mitigate side-channel attacks. However, the following algorithms were selected for comparison: HELO, AES, Blowfish, and Fernet, chosen for their effectiveness in resource-constrained settings (Table \ref{tab2}).

\begin{table}[htbp]
\caption{Parameters} 
\label{tab2}
\centering
\scriptsize 
\centering
\scalebox{1.5}{
\begin{tabular}{@{}lccccc@{}} 
\toprule
Algorithm & Key Size  & Block Size & Mode/Chunk & Padding/MAC & Rounds \\
\midrule
HELO      & 256 bits  & 512 bits   & Chunked Data & Poly1305    & 20     \\
AES       & 256 bits  & 128 bits   & CFB        & PKCS7       & 14     \\
Blowfish  & 256 bits  & 64 bits    & CFB        & PKCS7       & 16     \\
Fernet    & 256 bits  & 128 bits   & CBC        & PKCS7       & 10     \\
\bottomrule
\end{tabular}
}
\end{table}

The aspect of parameterization in encryption is essential as far as security, efficiency, and adaptation are concerned, particularly in environments like IoT devices and embedded systems, which are quite constrained. A key size plays an important role in defining the cryptographic strength of an algorithm, such that for brute-force attacks to be more resistant, the keys have to be larger \cite{kuwekar2014}. Furthermore, the bigger block size would improve the throughput, but it also affects the memory usage and latency \cite{parihar2016}. A particular encryption mode heavily affects the safety of data on the network because different modes, like Cipher Feedback (CFB) and Cipher Block Chaining (CBC) have different approaches to propagating data. They are vulnerable to replay, bit-flipping, and padding oracle \cite{Lu2021AES}. The use of padding schemes, like PKCS7, is in place to fit the data into a block cipher structure and prevent weaknesses related to such poorly structured data \cite{dijesh2020}. Furthermore, the most crucial parameter of an algorithm related to security strength is the number of encryption rounds. Most of the time, more rounds mean better protection against any differential and linear cryptanalysis; however, performance may significantly deteriorate as the number of rounds increases, especially in real-time and limited resources applications \cite{parihar2016}. Message Authentication Code (MAC), such as Poly1305 can also enhance the data integrity and authenticity of the encrypted form, such that any unauthorized modifications can be detected \cite{dijesh2020}. Also, these parameters determine the trade-offs between security and efficiency, favoring algorithm selection depending on the application's needs. Comparative performance evaluations have to take into consideration variations not only in comparison tests but also in the test configurations between successive tests. This is because variances in parameter settings can give rise to enormous potential differences in encryption speed, CPU utilization, memory consumption, and overall system efficiency \cite{kuwekar2014} \cite{Pronika2022FernetAES}. Lastly, by analyzing these parameters effectively, researchers and engineers can optimize their encryption solutions to achieve the most favorable balance between them and real-world applicability.

\subsubsection{CPU Processing Time Analysis}

\begin{figure}[htbp]
\centering
\begin{tikzpicture}
\begin{axis}[
    width=11cm,
    height=8.2cm,
    xlabel={File Size (B=Bytes, KB=Kilobytes)},
    ylabel={Time (Seconds)},
    title={CPU Processing Time for TXT Files},
    legend pos=north east,
    grid=major,
    symbolic x coords={15B, 118B, 171B, 1KB, 10KB, 22KB},
    xtick=data,
    ymin=0, ymax=0.13,
    ytick={0, 0.01, 0.02, 0.03, 0.04, 0.05, 0.06, 0.07, 0.08, 0.09, 0.1, 0.11, 0.12, 0.13},
    scaled ticks=false, 
    tick label style={/pgf/number format/fixed} 
]

\addplot[
    color=blue,
    mark=*,
    line width=1pt
] coordinates {
    (15B,0.01) (118B,0.01) (171B,0.03) (1KB,0.01) (10KB,0.03) (22KB,0.01)
};
\addlegendentry{HELO}

\addplot[
    color=red,
    mark=*,
    line width=1pt
] coordinates {
    (15B,0.03) (118B,0.03) (171B,0.05) (1KB,0.03) (10KB,0.02) (22KB,0.03)
};
\addlegendentry{AES}

\addplot[
    color=orange,
    mark=*,
    line width=1pt
] coordinates {
    (15B,0.05) (118B,0.09) (171B,0.06) (1KB,0.08) (10KB,0.08) (22KB,0.05)
};
\addlegendentry{Blowfish}

\addplot[
    color=teal,
    mark=*,
    line width=1pt
] coordinates {
    (15B,0.05) (118B,0.06) (171B,0.03) (1KB,0.06) (10KB,0.02) (22KB,0.03)
};
\addlegendentry{Fernet}

\end{axis}
\end{tikzpicture}
\caption{CPU Processing Time for TXT Files}
\label{graph2}
\end{figure}
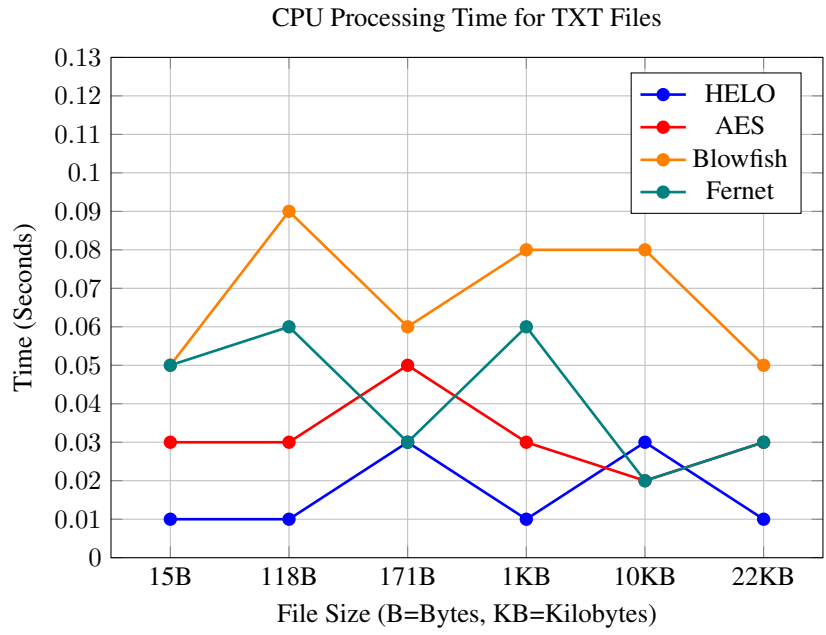

\textbf{TXT Files:}
The CPU processing time was analyzed across various file sizes, starting from small text files (15, 118, 171 bytes) to larger files (1, 10, 22 KB). The algorithms compared include HELO, AES, Blowfish, and Fernet. Our proposed model, HELO, consistently took 0.01 seconds for most file sizes, except for 171 bytes and 10 KB, where it recorded 0.03 seconds. Whereas AES showed varied CPU times between 0.02 to 0.05 seconds, while Blowfish had relatively higher processing times, ranging from 0.05 to 0.09 seconds without a clear pattern. Fernet had similarly inconsistent times, varying between 0.02 and 0.06 seconds. HELO averaged the lowest CPU time for TXT files at 0.016 seconds, followed by AES at 0.032 seconds, Fernet at 0.041 seconds, and Blowfish being the slowest with an average of 0.068 seconds (Figure \ref{graph2}).

\label{cpupro}

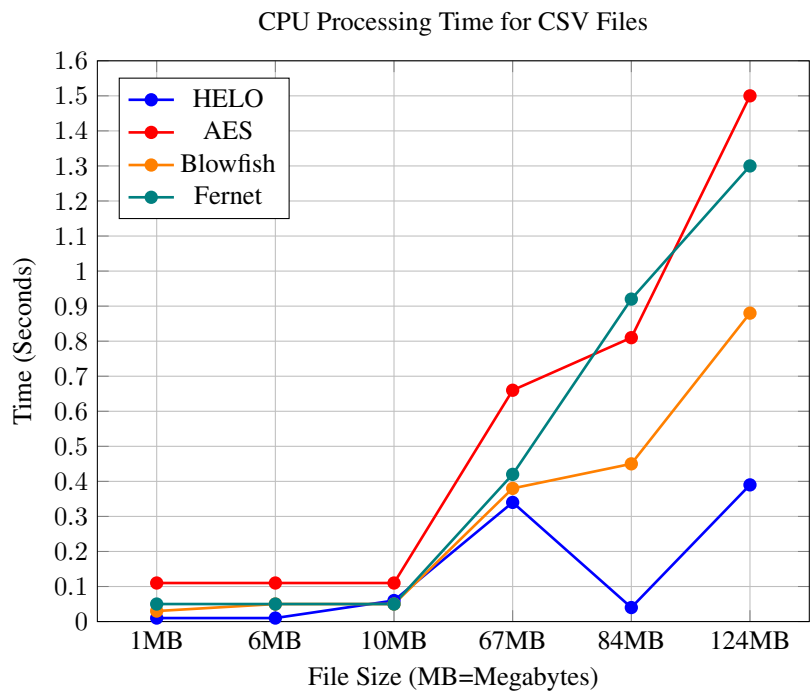
\begin{figure}[htbp]
\centering
\begin{tikzpicture}
\begin{axis}[
    width=11cm,
    height=9cm,    
    xlabel={File Size (MB=Megabytes)},
    ylabel={Time (Seconds)},
    title={CPU Processing Time for CSV Files},
    legend pos=north west,
    grid=major,
    symbolic x coords={1MB, 6MB, 10MB, 67MB, 84MB, 124MB},
    xtick=data,
    ymin=0, ymax=1.6,
    ytick={0, 0.1, 0.2, 0.3, 0.4, 0.5, 0.6, 0.7, 0.8, 0.9, 1.0, 1.1, 1.2, 1.3, 1.4, 1.5, 1.6},
    scaled ticks=false, 
    tick label style={/pgf/number format/fixed} 
]

\addplot[
    color=blue,
    mark=*,
    line width=1pt
] coordinates {
    (1MB,0.01) (6MB,0.01) (10MB,0.06) (67MB,0.34) (84MB,0.04) (124MB,0.39)
};
\addlegendentry{HELO}

\addplot[
    color=red,
    mark=*,
    line width=1pt
] coordinates {
    (1MB,0.11) (6MB,0.11) (10MB,0.11) (67MB,0.66) (84MB,0.81) (124MB,1.50)
};
\addlegendentry{AES}

\addplot[
    color=orange,
    mark=*,
    line width=1pt
] coordinates {
    (1MB,0.03) (6MB,0.05) (10MB,0.05) (67MB,0.38) (84MB,0.45) (124MB,0.88)
};
\addlegendentry{Blowfish}

\addplot[
    color=teal,
    mark=*,
    line width=1pt
] coordinates {
    (1MB,0.05) (6MB,0.05) (10MB,0.05) (67MB,0.42) (84MB,0.92) (124MB,1.30)
};
\addlegendentry{Fernet}

\end{axis}
\end{tikzpicture}
\caption{CPU Processing Time for CSV Files}
\label{graph3}
\end{figure}

\textbf{CSV Files:} HELO maintained its fast performance, processing small files (1 MB and 6 MB) in 0.01 seconds. The time increased with larger files, reaching 0.39 seconds for a 124 MB file, but still outperformed the other algorithms. AES exhibited steady growth in processing time, starting at 0.03 seconds for smaller files and increasing to 0.88 seconds for the largest. Blowfish maintained consistent times for smaller files but rose sharply to 1.5 seconds for the 124 MB file. Fernet followed a similar trend to Blowfish, processing smaller files efficiently but slowing down with larger ones. Overall, HELO was the fastest for CSV files, followed by AES and Fernet, while Blowfish took the longest (Figure \ref{graph3}).

\begin{figure}[htbp]
\centering
\begin{tikzpicture}
\begin{axis}[
    width=11cm,
    height=8cm,     
    xlabel={File Size (KB=Kilobytes, MB=Megabytes)},
    ylabel={Time (Seconds)},
    title={CPU Processing Time for Image Files},
    legend pos=north west,
    grid=major,
    symbolic x coords={368KB, 600KB, 740KB, 5MB, 20MB, 40MB},
    xtick=data,
    ymin=0, ymax=0.8,
    ytick={0, 0.1, 0.2, 0.3, 0.4, 0.5, 0.6, 0.7, 0.8},
    scaled ticks=false, 
    tick label style={/pgf/number format/fixed} 
]

\addplot[
    color=blue,
    mark=*,
    line width=1pt
] coordinates {
    (368KB,0.01) (600KB,0.01) (740KB,0.03) (5MB,0.15) (20MB,0.65) (40MB,0.75)
};
\addlegendentry{HELO}

\addplot[
    color=red,
    mark=*,
    line width=1pt
] coordinates {
    (368KB,0.06) (600KB,0.06) (740KB,0.05) (5MB,0.17) (20MB,0.31) (40MB,0.41)
};
\addlegendentry{AES}

\addplot[
    color=orange,
    mark=*,
    line width=1pt
] coordinates {
    (368KB,0.06) (600KB,0.06) (740KB,0.05) (5MB,0.14) (20MB,0.34) (40MB,0.41)
};
\addlegendentry{Blowfish}

\addplot[
    color=teal,
    mark=*,
    line width=1pt
] coordinates {
    (368KB,0.03) (600KB,0.02) (740KB,0.03) (5MB,0.12) (20MB,0.25) (40MB,0.22)
};
\addlegendentry{Fernet}

\end{axis}
\end{tikzpicture}
\caption{CPU Processing Time for Image Files}
\label{graph4}
\end{figure}
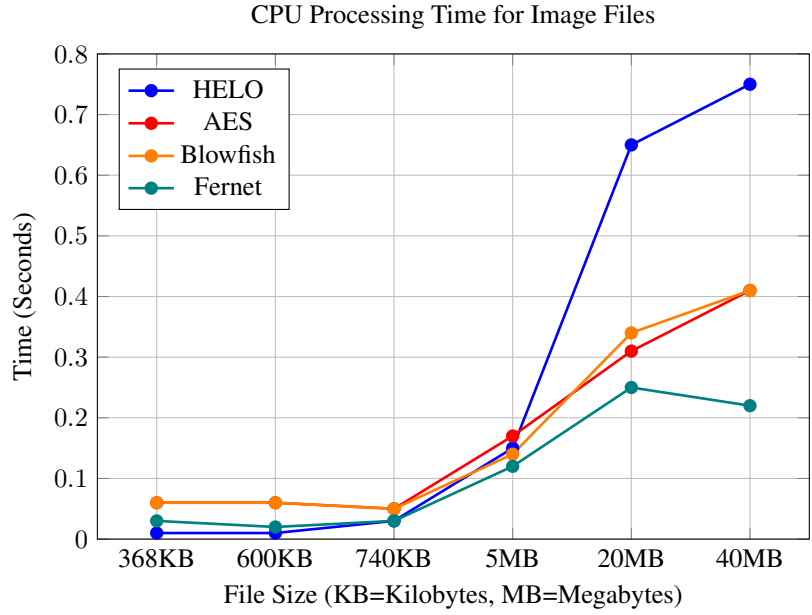

\textbf{Image Files:} For image files, the results were different. HELO performed best for smaller images (368 KB to 740 KB) with times ranging from 0.01 to 0.03 seconds. However, as file sizes increased to 5 MB, 20 MB, and 40 MB, HELO’s time rose to 0.75 seconds for the 40 MB file, making it slower than AES (0.14 seconds), Blowfish (0.41 seconds), and Fernet (0.22 seconds). HELO’s use of multithreading, while usually beneficial, may have caused overhead when processing larger image files, leading to reduced efficiency. Therefore, HELO was the fastest for smaller image files, but AES was the best for larger ones (Figure \ref{graph4}).

\subsubsection{RAM Usage Analysis}
\begin{figure}[htbp]
\centering
\begin{tikzpicture}
\begin{axis}[
    width=11cm,
    height=9cm,       
    xlabel={File Size (B=Bytes, KB=Kilobytes)},
    ylabel={RAM Usage (Megabytes)},
    title={RAM Usage for TXT Files},
    legend pos=south east,
    grid=major,
    symbolic x coords={15B, 118B, 171B, 1KB, 10KB, 22KB},
    xtick=data,
    ymin=0, ymax=10,
    ytick={0, 1.0, 2.0, 3.0, 4.0, 5.0, 6.0, 7.0, 8.0, 9.0, 10},
    scaled ticks=false, 
    tick label style={/pgf/number format/fixed} 
]

\addplot[
    color=blue,
    mark=*,
    line width=1pt
] coordinates {
    (15B,4.74) (118B,4.87) (171B,4.78) (1KB,4.87) (10KB,4.96) (22KB,4.81)
};
\addlegendentry{HELO}

\addplot[
    color=red,
    mark=*,
    line width=1pt
] coordinates {
    (15B,6.41) (118B,6.51) (171B,6.25) (1KB,6.42) (10KB,6.23) (22KB,6.50)
};
\addlegendentry{AES}

\addplot[
    color=orange,
    mark=*,
    line width=1pt
] coordinates {
    (15B,6.37) (118B,6.30) (171B,6.38) (1KB,6.31) (10KB,6.38) (22KB,6.60)
};
\addlegendentry{Blowfish}

\addplot[
    color=teal,
    mark=*,
    line width=1pt
] coordinates {
    (15B,6.05) (118B,6.41) (171B,6.04) (1KB,6.12) (10KB,6.26) (22KB,6.15)
};
\addlegendentry{Fernet}

\end{axis}
\end{tikzpicture}
\caption{RAM Usage for TXT Files}
\label{gra5}
\end{figure}
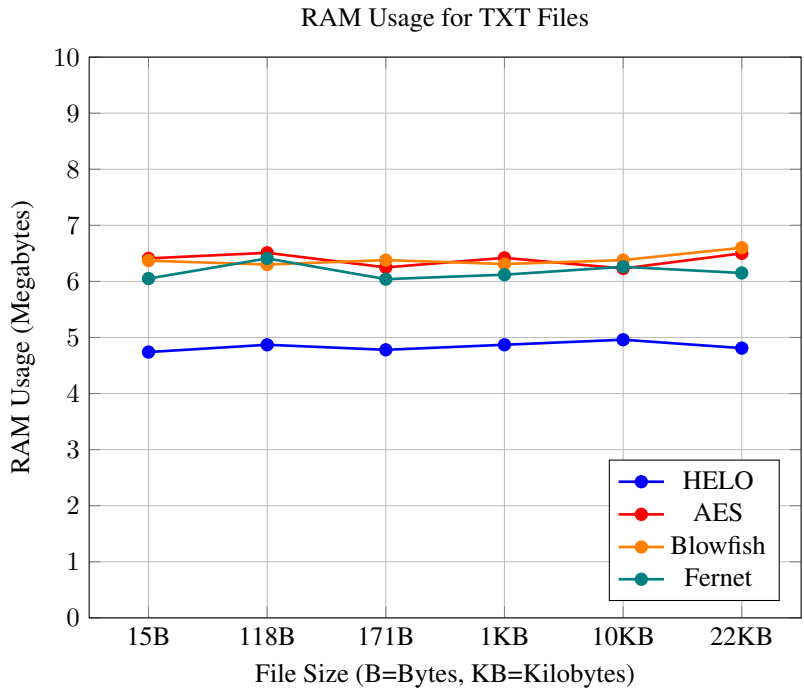

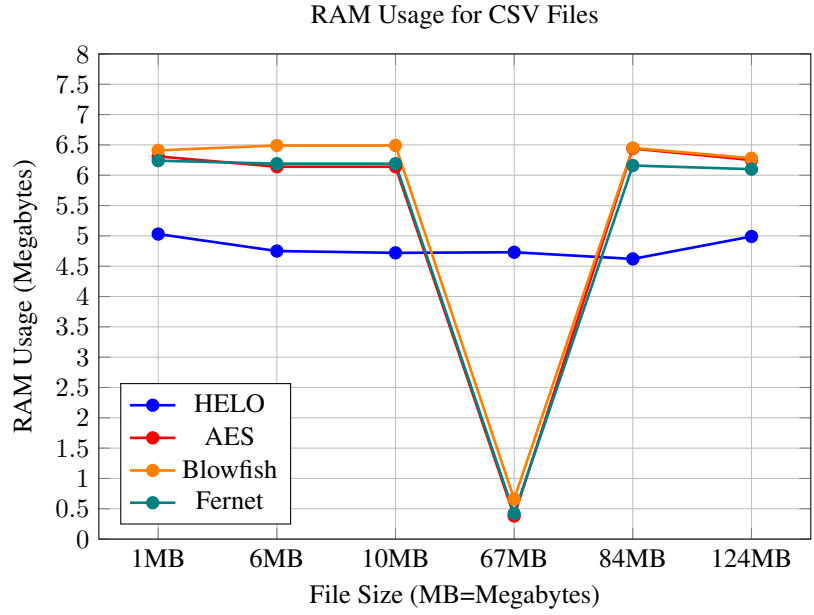
\begin{figure}[htbp]
\centering
\begin{tikzpicture}
\begin{axis}[
    width=11cm,
    height=8cm,            
    xlabel={File Size (MB=Megabytes)},
    ylabel={RAM Usage (Megabytes)},
    title={RAM Usage for CSV Files},
    legend pos=south west,
    grid=major,
    symbolic x coords={1MB, 6MB, 10MB, 67MB, 84MB, 124MB},
    xtick=data,
    ymin=0, ymax=8,
    ytick={0, 0.5, 1, 1.5, 2, 2.5, 3, 3.5, 4, 4.5, 5, 5.5, 6, 6.5, 7, 7.5, 8},
    scaled ticks=false, 
    tick label style={/pgf/number format/fixed} 
]

\addplot[
    color=blue,
    mark=*,
    line width=1pt
] coordinates {
    (1MB,5.03) (6MB,4.75) (10MB,4.72) (67MB,4.73) (84MB,4.62) (124MB,4.99)
};
\addlegendentry{HELO}

\addplot[
    color=red,
    mark=*,
    line width=1pt
] coordinates {
    (1MB,6.31) (6MB,6.14) (10MB,6.14) (67MB,0.38) (84MB,6.44) (124MB,6.25)
};
\addlegendentry{AES}

\addplot[
    color=orange,
    mark=*,
    line width=1pt
] coordinates {
    (1MB,6.41) (6MB,6.49) (10MB,6.49) (67MB,0.66) (84MB,6.45) (124MB,6.28)
};
\addlegendentry{Blowfish}

\addplot[
    color=teal,
    mark=*,
    line width=1pt
] coordinates {
    (1MB,6.24) (6MB,6.19) (10MB,6.19) (67MB,0.42) (84MB,6.16) (124MB,6.10)
};
\addlegendentry{Fernet}

\end{axis}
\end{tikzpicture}
\caption{RAM Usage for CSV Files}
\label{gra6}
\end{figure}

\textbf{TXT and CSV Files:} The TXT files (Figure \ref{gra5}) and CSV files (Figure \ref{gra6}) showed similar trends across algorithms with HELO, using approximately 4.70 MB of RAM, AES around 6.30 MB, Blowfish at 6.40 MB, and Fernet using 6.23 MB. As a result, it proves that HELO’s lower RAM usage for TXT and CSV files indicates better memory efficiency. This analysis shows that HELO consistently showed lower RAM usage than the other algorithms.

\begin{figure}[H]
\centering
\begin{tikzpicture}
\begin{axis}[
    width=11cm,
    height=9cm,            
    xlabel={File Size (KB=Kilobytes, MB=Megabytes)},
    ylabel={RAM Usage (Megabytes)},
    title={RAM Usage for Image Files},
    legend pos=south east,
    grid=major,
    symbolic x coords={368KB, 600KB, 740KB, 5MB, 20MB, 40MB},
    xtick=data,
    ymin=0, ymax=10.0,
    ytick={0, 0.5, 1.0, 1.5, 2.0, 2.5, 3.0, 3.5, 4.0, 4.5, 5.0, 5.5, 6.0, 6.5, 7.0, 7.5, 8.0, 8.5, 9.0, 9.5, 10.0},
    scaled ticks=false, 
    tick label style={/pgf/number format/fixed} 
]

\addplot[
    color=blue,
    mark=*,
    line width=1pt
] coordinates {
    (368KB,5.31) (600KB,5.11) (740KB,4.80) (5MB,7.98) (20MB,9.17) (40MB,7.93)
};
\addlegendentry{HELO}

\addplot[
    color=red,
    mark=*,
    line width=1pt
] coordinates {
    (368KB,6.46) (600KB,7.05) (740KB,6.55) (5MB,7.84) (20MB,8.87) (40MB,9.75)
};
\addlegendentry{AES}

\addplot[
    color=orange,
    mark=*,
    line width=1pt
] coordinates {
    (368KB,6.66) (600KB,7.01) (740KB,6.20) (5MB,7.72) (20MB,8.17) (40MB,8.92)
};
\addlegendentry{Blowfish}

\addplot[
    color=teal,
    mark=*,
    line width=1pt
] coordinates {
    (368KB,6.80) (600KB,6.96) (740KB,6.07) (5MB,8.09) (20MB,7.92) (40MB,9.36)
};
\addlegendentry{Fernet}

\end{axis}
\end{tikzpicture}
\caption{RAM Usage for Image Files}
\label{gra7}
\end{figure}
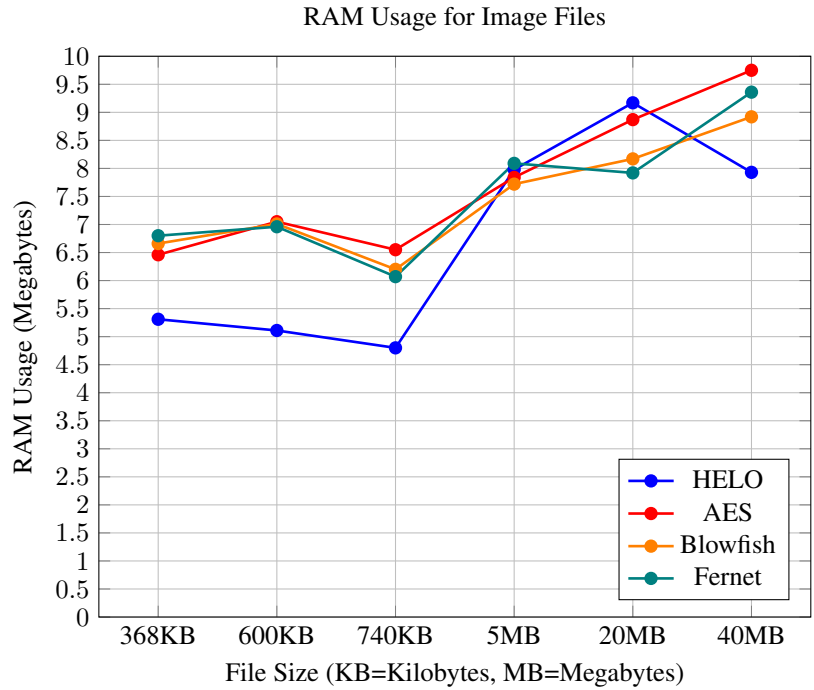

\textbf{Image Files:} For image files, as the file sizes grew, HELO's RAM usage increased, peaking at 9.17 MB for a 20 MB file, before dropping to 7.93 MB for a 40 MB file. Whereas, the memory usage of the other algorithms like AES, Fernet, and Blowfish has a peak value of 9.8 MB, 9.4 MB, and 8.9 MB respectively. In contrast, AES, Blowfish, and Fernet had more consistent RAM usage for larger files compared to HELO (Figure \ref{gra7}).

\subsubsection{Runtime Analysis}
\label{runtime}
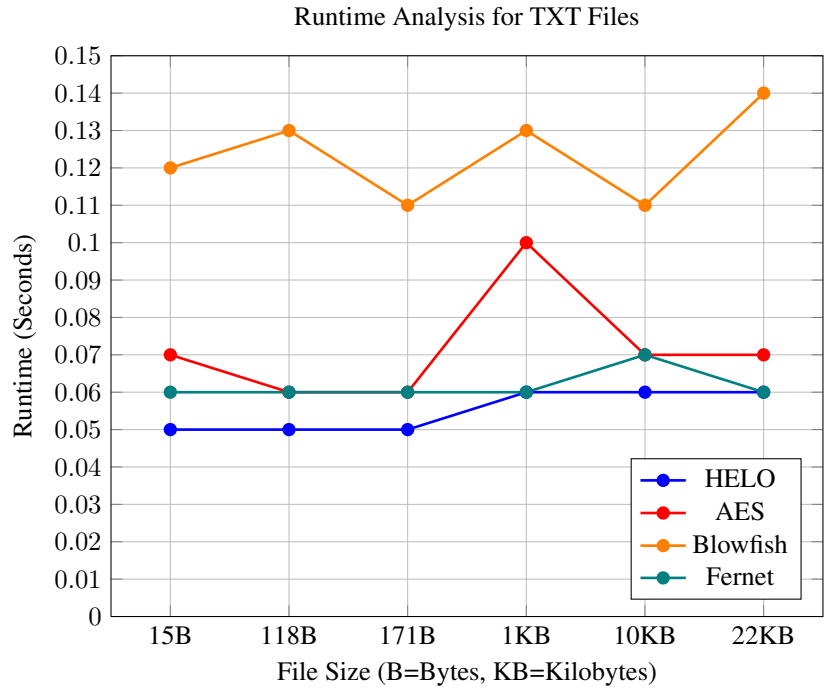
\begin{figure}[htbp]
\centering
\begin{tikzpicture}
\begin{axis}[
    width=11cm,
    height=9cm,         
    xlabel={File Size (B=Bytes, KB=Kilobytes)},
    ylabel={Runtime (Seconds)},
    title={Runtime Analysis for TXT Files},
    legend pos=south east,
    grid=major,
    symbolic x coords={15B, 118B, 171B, 1KB, 10KB, 22KB},
    xtick=data,
    ymin=0, ymax=0.15,
    ytick={0, 0.01, 0.02, 0.03, 0.04, 0.05, 0.06, 0.07, 0.08, 0.09, 0.10, 0.11, 0.12, 0.13, 0.14, 0.15},
    scaled ticks=false, 
    tick label style={/pgf/number format/fixed} 
]

\addplot[
    color=blue,
    mark=*,
    line width=1pt
] coordinates {
    (15B,0.05) (118B,0.05) (171B,0.05) (1KB,0.06) (10KB,0.06) (22KB,0.06)
};
\addlegendentry{HELO}

\addplot[
    color=red,
    mark=*,
    line width=1pt
] coordinates {
    (15B,0.07) (118B,0.06) (171B,0.06) (1KB,0.1) (10KB,0.07) (22KB,0.07)
};
\addlegendentry{AES}

\addplot[
    color=orange,
    mark=*,
    line width=1pt
] coordinates {
    (15B,0.12) (118B,0.13) (171B,0.11) (1KB,0.13) (10KB,0.11) (22KB,0.14)
};
\addlegendentry{Blowfish}

\addplot[
    color=teal,
    mark=*,
    line width=1pt
] coordinates {
    (15B,0.06) (118B,0.06) (171B,0.06) (1KB,0.06) (10KB,0.07) (22KB,0.06)
};
\addlegendentry{Fernet}

\end{axis}
\end{tikzpicture}
\caption{Runtime Analysis for TXT Files}
\label{gra8}
\end{figure}

The following figures are generated for Runtime Analysis. These figures will be used to analyze each type of file for different types of algorithms based on the runtime they have had. Runtime is the amount of time a computer program or algorithm takes to execute a program. In this case, runtime represents how long it takes for the encryption or decryption process to complete for a given file size.\\
\\
\textbf{TXT Files:} For TXT files, the runtime analysis revealed that our proposed cryptographic model consistently had the fastest performance. It recorded an average runtime of 0.05 seconds, outperforming the other algorithms by a significant margin. Whereas, Fernet followed closely behind, showing comparable efficiency, but not quite matching HELO's speed. In addition, AES demonstrated moderate performance with an average runtime of 0.07 seconds, slightly slower than Fernet and HELO. However, Blowfish was the slowest algorithm when processing the TXT files, recording an average runtime of 0.11 seconds, making it almost twice as slow as HELO cryptography. As a result, this analysis suggests that HELO is particularly well-suited for handling smaller, less complex file types like TXT, where the processing demands are relatively low and speed is paramount (Figure \ref{gra8}).

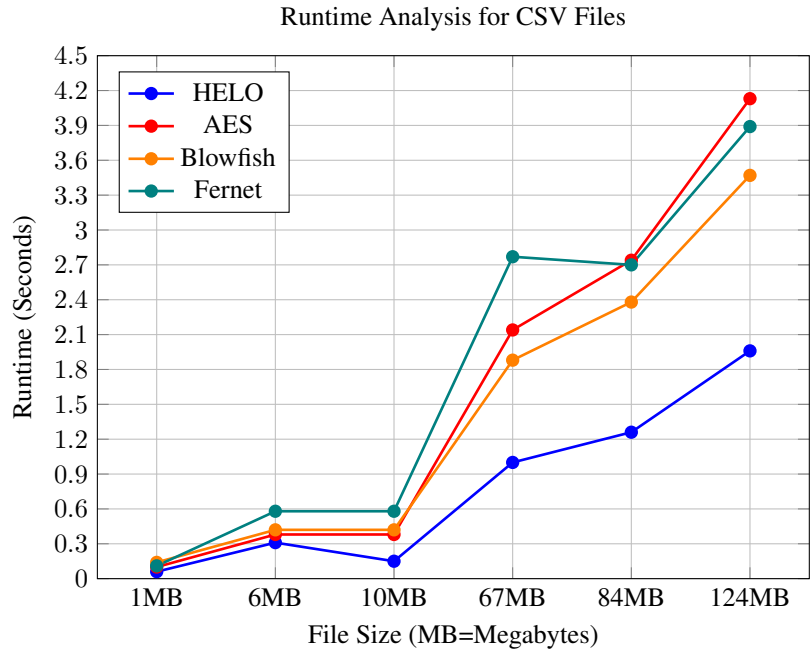
\begin{figure}[htbp]
\centering
\begin{tikzpicture}
\begin{axis}[
    width=11cm,
    height=8.5cm,            
    xlabel={File Size (MB=Megabytes)},
    ylabel={Runtime (Seconds)},
    title={Runtime Analysis for CSV Files},
    legend pos=north west,
    grid=major,
    symbolic x coords={1MB, 6MB, 10MB, 67MB, 84MB, 124MB},
    xtick=data,
    ymin=0, ymax=4.5,
    ytick={0, 0.3, 0.6, 0.9, 1.2, 1.5, 1.8, 2.1, 2.4, 2.7, 3.0, 3.3, 3.6, 3.9, 4.2, 4.5},
    scaled ticks=false, 
    tick label style={/pgf/number format/fixed} 
]

\addplot[
    color=blue,
    mark=*,
    line width=1pt
] coordinates {
    (1MB,0.06) (6MB,0.31) (10MB,0.15) (67MB,1.00) (84MB,1.26) (124MB,1.96)
};
\addlegendentry{HELO}

\addplot[
    color=red,
    mark=*,
    line width=1pt
] coordinates {
    (1MB,0.10) (6MB,0.38) (10MB,0.38) (67MB,2.14) (84MB,2.74) (124MB,4.13)
};
\addlegendentry{AES}

\addplot[
    color=orange,
    mark=*,
    line width=1pt
] coordinates {
    (1MB,0.14) (6MB,0.42) (10MB,0.42) (67MB,1.88) (84MB,2.38) (124MB,3.47)
};
\addlegendentry{Blowfish}

\addplot[
    color=teal,
    mark=*,
    line width=1pt
] coordinates {
    (1MB,0.11) (6MB,0.58) (10MB,0.58) (67MB,2.77) (84MB,2.70) (124MB,3.89)
};
\addlegendentry{Fernet}

\end{axis}
\end{tikzpicture}
\caption{Runtime Analysis for CSV Files}
\label{gra9}
\end{figure}

\begin{figure}[htbp]
\centering
\begin{tikzpicture}
\begin{axis}[
    width=11cm,
    height=8.5cm,            
    xlabel={File Size (KB=Kilobytes, MB=Megabytes)},
    ylabel={Runtime (Seconds)},
    title={Runtime Analysis for Image Files},
    legend pos=north west,
    grid=major,
    symbolic x coords={368KB, 600KB, 740KB, 5MB, 20MB, 40MB},
    xtick=data,
    ymin=0, ymax=8.5,
    ytick={0, 0.5, 1.0, 1.5, 2.0, 2.5, 3.0, 3.5, 4.0, 4.5, 5.0, 5.5, 6.0, 6.5, 7.0, 7.5, 8.0, 8.5},
    scaled ticks=false, 
    tick label style={/pgf/number format/fixed} 
]

\addplot[
    color=blue,
    mark=*,
    line width=1pt
] coordinates {
    (368KB,0.15) (600KB,0.09) (740KB,0.07) (5MB,2.60) (20MB,3.88) (40MB,3.22)
};
\addlegendentry{HELO}

\addplot[
    color=red,
    mark=*,
    line width=1pt
] coordinates {
    (368KB,0.53) (600KB,0.19) (740KB,0.12) (5MB,3.53) (20MB,6.91) (40MB,7.91)
};
\addlegendentry{AES}

\addplot[
    color=orange,
    mark=*,
    line width=1pt
] coordinates {
    (368KB,0.58) (600KB,0.24) (740KB,0.15) (5MB,3.39) (20MB,6.58) (40MB,4.73)
};
\addlegendentry{Blowfish}

\addplot[
    color=teal,
    mark=*,
    line width=1pt
] coordinates {
    (368KB,0.63) (600KB,0.58) (740KB,0.11) (5MB,3.44) (20MB,5.67) (40MB,5.29)
};
\addlegendentry{Fernet}

\end{axis}
\end{tikzpicture}
\caption{Runtime Analysis for Image Files}
\label{gra10}
\end{figure}
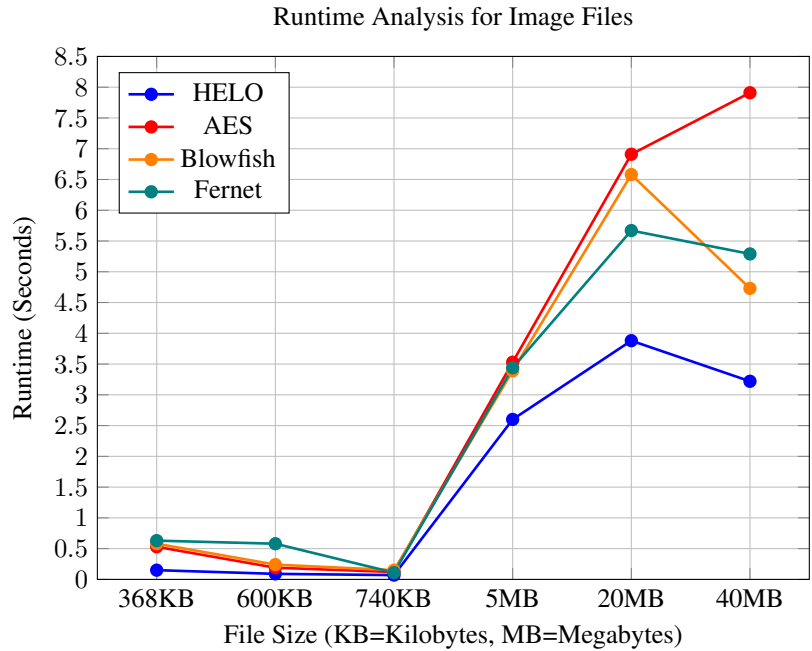

\textbf{CSV Files:} In the case of CSV files, our proposed cryptographic model, HELO again maintained its position as the fastest algorithm. For smaller file sizes, such as the 1 MB CSV file, HELO processed it in just 0.06 seconds, demonstrating its efficiency in handling structured data files. However, as the file sizes grew, the runtime naturally increased, with HELO taking 1.96 seconds to process a 124 MB CSV file. Even with this increase, HELO still outperformed the other algorithms. However, AES and Blowfish exhibited significantly higher runtimes, especially as the file sizes grew. Therefore, this pattern highlights HELO's ability to manage structured data effectively, making it the most efficient choice for CSV file encryption and decryption tasks. While AES and Blowfish are capable of processing such files, their runtimes increased more sharply with the file size, indicating that they are struggling with large datasets compared to HELO cryptography (Figure \ref{gra9}).\\
\\
\textbf{Image Files:} For image files, the runtime analysis took a different turn. While HELO was still fast for smaller image files, such as those around 368 KB to 740 KB, where it took between 0.01 to 0.03 seconds, its performance began to taper off with larger files. When processing a 40 MB image file, HELO’s runtime rose significantly to 0.75 seconds. This was slower than AES, which recorded a runtime of 0.14 seconds for the same file, making AES the better choice for larger image files. Blowfish, with a runtime of 0.41 seconds, and Fernet, with a runtime of 0.22 seconds, also performed better than HELO for larger images. The results suggest that HELO's efficiency diminishes as file size increases, especially for more complex files like images. This is likely due to the overhead created by HELO’s multithreading process, which, while beneficial for smaller and simpler files, may become a bottleneck for larger and more complex file types. Therefore, AES emerged as the most suitable algorithm for handling large image files, demonstrating faster runtimes and more consistent performance (Figure \ref{gra10}).

\subsection{Energy Consumption Analysis}
The results of the CPU processing time from our performance analysis will be used to measure the energy consumption of each cryptographic algorithm. Here, the unit of CPU processing time is seconds, watts for power, and joules for energy. In addition, the formula used to analyze energy consumption is provided below.\\
\\
\(
\begin{aligned}
Energy = Power \times Time
\end{aligned}
\)
\\
\\
\textbf{TXT Files:} Results from energy consumption for the TXT files vary across several file sizes. Each cryptographic algorithm produces different results based on the time taken and the power used during the process. Firstly, HELO consumed 0.95 Joules of energy for 1 Byte of file, and this remained constant for 118 Bytes, 1 KB, and 22 KB. However, energy consumption slightly increased to 2.85 Joules for 171 KB and 10 KB. In contrast, the other algorithms used more energy to operate efficiently. In this analysis, AES required 3 times more energy than HELO for 1 Byte, 118 Bytes, 1 KB, and 22 KB. It used 1.67 times more energy for 171 Bytes and 33.33\% less energy than HELO for 10 KB. Likewise, Blowfish consumed 5 times more energy for 1 Byte and 22 KB, 9 times more for 118 Bytes, 2 times more for 171 Bytes, 8 times more for 1 KB, and 2.67 times more for 10 KB than HELO. Similarly, Fernet utilized 5 times more energy for 1 Byte, 6 times more for 118 Bytes and 1 KB, and 3 times more for 22 Bytes than HELO. However, both HELO and Fernet consumed equal energy for 171 Bytes and 10 KB. After a competitive analysis, it is evident that HELO surpasses other algorithms in terms of consuming less energy to execute efficiently (Table \ref{tab21}).

\begin{table}[htbp]
\centering
\caption{Energy Consumption for TXT Files}
\label{tab21}%
\scalebox{0.93}{
\begin{tabular}{|p{2cm}|p{3cm}|p{2.5cm}|p{2cm}|p{2.2cm}|}
\hline
\textbf{File Size} & \textbf{Cryptography} & \textbf{Power (W)} & \textbf{Time (S)} & \textbf{Energy (J)} \\
\hline
\multirow{4}{*}{1 Byte} 
& HELO  & \multirow{4}{*}{95} & 0.01 & 0.95 \\
\cline{2-2} \cline{4-5}
& AES &                       & 0.03 & 2.85 \\
\cline{2-2} \cline{4-5}
& Blowfish &                  & 0.05 & 4.75 \\
\cline{2-2} \cline{4-5}
& Fernet &                    & 0.05 & 4.75 \\
\hline
\multirow{4}{*}{118 Bytes}
& HELO  & \multirow{4}{*}{95} & 0.01 & 0.95 \\
\cline{2-2} \cline{4-5}
& AES &                       & 0.03 & 2.85 \\
\cline{2-2} \cline{4-5}
& Blowfish &                  & 0.09 & 8.55 \\
\cline{2-2} \cline{4-5}
& Fernet &                    & 0.06 & 5.7 \\
\hline
\multirow{4}{*}{171 Bytes}
& HELO  & \multirow{4}{*}{95} & 0.03 & 2.85 \\
\cline{2-2} \cline{4-5}
& AES &                       & 0.05 & 4.75 \\
\cline{2-2} \cline{4-5}
& Blowfish &                  & 0.06 & 5.7 \\
\cline{2-2} \cline{4-5}
& Fernet &                    & 0.03 & 2.85 \\
\hline
\multirow{4}{*}{1 KB}
& HELO  & \multirow{4}{*}{95} & 0.01 & 0.95 \\
\cline{2-2} \cline{4-5}
& AES &                       & 0.03 & 2.85 \\
\cline{2-2} \cline{4-5}
& Blowfish &                  & 0.08 & 7.6 \\
\cline{2-2} \cline{4-5}
& Fernet &                    & 0.06 & 5.7 \\
\hline
\multirow{4}{*}{10 KB}
& HELO  & \multirow{4}{*}{95} & 0.03 & 2.85 \\
\cline{2-2} \cline{4-5}
& AES &                       & 0.02 & 1.9 \\
\cline{2-2} \cline{4-5}
& Blowfish &                  & 0.08 & 7.6 \\
\cline{2-2} \cline{4-5}
& Fernet &                    & 0.03 & 2.85 \\
\hline
\multirow{4}{*}{22 KB}
& HELO  & \multirow{4}{*}{95} & 0.01 & 0.95 \\
\cline{2-2} \cline{4-5}
& AES &                       & 0.03 & 2.85 \\
\cline{2-2} \cline{4-5}
& Blowfish &                  & 0.05 & 4.75 \\
\cline{2-2} \cline{4-5}
& Fernet &                    & 0.03 & 2.85 \\
\hline
\end{tabular}
}
\end{table}

\textbf{CSV Files:} For CSV files, AES required 10 times more energy than HELO for 1 MB and 6 MB, 1.67 times more for 10 MB, 2.03 times more for 67 MB, 20.25 times more for 84 MB, and 3.85 times more for 124 MB. Additionally, Blowfish consumed 3 times more energy for 1 MB, 5 times more for 6 MB, 1.81 times more for 67 MB, 11.25 times more for 84 MB, and 2.28 times more for 124 MB than HELO. Similarly, Fernet utilized 5 times more energy for 1 MB, 5 times more for 6 MB, 1.27 times more for 67 MB, 23 times more for 84 MB, and 3.33 times more for 124 MB than HELO. Furthermore, Blowfish and Fernet both consumed 5 and 1.2 times more energy than HELO for 6 MB and 10 MB of a CSV file. Finally, this analysis also indicates that HELO consumes significantly less energy compared to the other cryptographic algorithms (Table \ref{tab22}).\\
\\
\textbf{Image Files:} For image files, both AES and Blowfish required 6 times more energy than HELO for 368 KB and 600 KB. They also consumed 1.67 times more energy for a 740 KB file. Additionally, Fernet used 3 times more energy for 368 KB and 2 times more for 600 KB of image files. However, HELO struggled with performance when handling larger files. Consequently, the CPU required significantly more time to operate, which resulted in higher energy consumption compared to the other algorithms. According to the analysis, AES consumed 52.31\% less energy at 20 MB. Furthermore, Blowfish and Fernet used 47.69\% and 61.54\% less energy, respectively, for the same file size. Similarly, AES and Blowfish utilized 45.33\% less energy at 40 MB, while Fernet needed 69.26\% less energy than HELO. This analysis indicates that HELO is very effective for smaller image files but degrades its efficiency in energy consumption with larger ones (Table \ref{tab23}).

\begin{table}[htbp]
\caption{Energy Consumption for CSV Files}
\label{tab22}%
\centering
\scalebox{0.93}{
\begin{tabular}{|p{2cm}|p{3cm}|p{2.5cm}|p{2cm}|p{2.2cm}|}
\hline
\textbf{File Size} & \textbf{Cryptography} & \textbf{Power (W)} & \textbf{Time (S)} & \textbf{Energy (J)} \\
\hline
\multirow{4}{*}{1 MB} 
& HELO  & \multirow{4}{*}{95} & 0.01 & 0.95 \\
\cline{2-2} \cline{4-5}
& AES &                       & 0.1 & 9.5 \\
\cline{2-2} \cline{4-5}
& Blowfish &                  & 0.03 & 2.85 \\
\cline{2-2} \cline{4-5}
& Fernet &                    & 0.05 & 4.75 \\
\hline
\multirow{4}{*}{6 MB}
& HELO  & \multirow{4}{*}{95} & 0.01 & 0.95 \\
\cline{2-2} \cline{4-5}
& AES &                       & 0.1 & 9.5 \\
\cline{2-2} \cline{4-5}
& Blowfish &                  & 0.05 & 4.75 \\
\cline{2-2} \cline{4-5}
& Fernet &                    & 0.05 & 4.75 \\
\hline
\multirow{4}{*}{10 MB}
& HELO  & \multirow{4}{*}{95} & 0.06 & 5.7 \\
\cline{2-2} \cline{4-5}
& AES &                       & 0.1 & 9.5 \\
\cline{2-2} \cline{4-5}
& Blowfish &                  & 0.05 & 4.75 \\
\cline{2-2} \cline{4-5}
& Fernet &                    & 0.05 & 4.75 \\
\hline
\multirow{4}{*}{67 MB}
& HELO  & \multirow{4}{*}{95} & 0.33 & 31.35 \\
\cline{2-2} \cline{4-5}
& AES &                       & 0.67 & 63.65 \\
\cline{2-2} \cline{4-5}
& Blowfish &                  & 0.39 & 37.05 \\
\cline{2-2} \cline{4-5}
& Fernet &                    & 0.42 & 39.9 \\
\hline
\multirow{4}{*}{84 MB}
& HELO  & \multirow{4}{*}{95} & 0.04 & 3.8 \\
\cline{2-2} \cline{4-5}
& AES &                       & 0.81 & 76.95 \\
\cline{2-2} \cline{4-5}
& Blowfish &                  & 0.45 & 42.75 \\
\cline{2-2} \cline{4-5}
& Fernet &                    & 0.92 & 87.4 \\
\hline
\multirow{4}{*}{124 MB}
& HELO  & \multirow{4}{*}{95} & 0.39 & 37.05 \\
\cline{2-2} \cline{4-5}
& AES &                       & 1.5 & 142.5 \\
\cline{2-2} \cline{4-5}
& Blowfish &                  & 0.89 & 84.55 \\
\cline{2-2} \cline{4-5}
& Fernet &                    & 1.3 & 123.5 \\
\hline
\end{tabular}
}
\end{table}

\begin{table}[htbp]
\caption{Energy Consumption for Image Files}
\label{tab23}%
\centering
\scalebox{0.93}{
\begin{tabular}{|p{2cm}|p{3cm}|p{2.5cm}|p{2cm}|p{2.2cm}|}
\hline
\textbf{File Size} & \textbf{Cryptography} & \textbf{Power (W)} & \textbf{Time (S)} & \textbf{Energy (J)} \\
\hline
\multirow{4}{*}{368 KB} 
& HELO  & \multirow{4}{*}{95} & 0.01 & 0.95 \\
\cline{2-2} \cline{4-5}
& AES &                       & 0.06 & 5.7 \\
\cline{2-2} \cline{4-5}
& Blowfish &                  & 0.06 & 5.7 \\
\cline{2-2} \cline{4-5}
& Fernet &                    & 0.03 & 2.85 \\
\hline
\multirow{4}{*}{600 KB}
& HELO  & \multirow{4}{*}{95} & 0.01 & 0.95 \\
\cline{2-2} \cline{4-5}
& AES &                       & 0.06 & 5.7 \\
\cline{2-2} \cline{4-5}
& Blowfish &                  & 0.06 & 5.7 \\
\cline{2-2} \cline{4-5}
& Fernet &                    & 0.02 & 1.9 \\
\hline
\multirow{4}{*}{740 KB}
& HELO  & \multirow{4}{*}{95} & 0.03 & 2.85 \\
\cline{2-2} \cline{4-5}
& AES &                       & 0.05 & 4.75 \\
\cline{2-2} \cline{4-5}
& Blowfish &                  & 0.05 & 4.75 \\
\cline{2-2} \cline{4-5}
& Fernet &                    & 0.03 & 2.85 \\
\hline
\multirow{4}{*}{5 MB}
& HELO  & \multirow{4}{*}{95} & 0.15 & 14.25 \\
\cline{2-2} \cline{4-5}
& AES &                       & 0.17 & 16.15 \\
\cline{2-2} \cline{4-5}
& Blowfish &                  & 0.14 & 13.3 \\
\cline{2-2} \cline{4-5}
& Fernet &                    & 0.12 & 11.4 \\
\hline
\multirow{4}{*}{20 MB}
& HELO  & \multirow{4}{*}{95} & 0.65 & 61.75 \\
\cline{2-2} \cline{4-5}
& AES &                       & 0.31 & 29.45 \\
\cline{2-2} \cline{4-5}
& Blowfish &                  & 0.34 & 32.3 \\
\cline{2-2} \cline{4-5}
& Fernet &                    & 0.25 & 23.75 \\
\hline
\multirow{4}{*}{40 MB}
& HELO  & \multirow{4}{*}{95} & 0.75 & 71.25 \\
\cline{2-2} \cline{4-5}
& AES &                       & 0.41 & 38.95 \\
\cline{2-2} \cline{4-5}
& Blowfish &                  & 0.41 & 38.95 \\
\cline{2-2} \cline{4-5}
& Fernet &                    & 0.22 & 20.9 \\
\hline
\end{tabular}
}
\end{table}

\subsection{Avalanche Effect Analysis}
Cryptography uses multiple methods to protect data. One of the most important ones is the avalanche effect. It promises that even a small change in the input will result in a significant change in the output. Also, this analysis enhances security by making it critical for unauthorized users to decode the original data without the correct decryption key. The formula for calculating the average value of the avalanche effect of HELO corresponding to bit positions is given below.\\
\\
\(
\begin{aligned}
\text{Average Avalanche Effect (\%)} = \frac{1}{T} \left( \sum_{i=0}^{n} C_i \right)
\end{aligned}
\)\\
\\
\\
Here, \(T\) is the total number of bit positions in the ciphertext which is 21, \(i\) is the initial index of the flipping bit which is 0, \(n\) is the total number of bits flipped which is 20, and \(C_i\) is the value of the avalanche effect in each bit position. Consequently, the value of \(T\) will be \(n+1\).

\begin{table}[htbp]
\caption{Avalanche Effect Analysis of TXT Files}
\label{tab12}%
\scriptsize 
\centering
\scalebox{1.25}{
\begin{tabular}{@{}lllllll@{}}
\toprule
Bit position & 15 B & 118 B & 171 B & 1 KB & 10 KB & 22 KB \\
\midrule
0           & 59.17 & 52.97  & 50.73 & 50.75 & 50.17 & 50.10 \\
1           & 50.00 & 47.88  & 49.64 & 49.10 & 49.72 & 49.45 \\
2           & 47.50 & 50.11  & 50.80 & 50.42 & 49.83 & 49.79 \\
3           & 46.67 & 50.00  & 46.27 & 50.63 & 49.75 & 49.91 \\
4           & 50.83 & 52.01  & 48.46 & 50.05 & 49.94 & 49.44 \\
5           & 55.00 & 47.56  & 51.02 & 50.64 & 49.87 & 50.37 \\
6           & 45.00 & 49.68  & 49.20 & 50.29 & 49.89 & 50.15 \\
7           & 50.00 & 49.89  & 48.31 & 49.00 & 49.94 & 49.88 \\
8           & 45.83 & 47.35  & 53.51 & 49.28 & 49.85 & 50.49 \\
9           & 45.83 & 50.42  & 50.80 & 49.64 & 49.93 & 49.41 \\
10          & 59.17 & 51.69  & 51.38 & 50.01 & 49.84 & 50.08 \\
11          & 47.50 & 49.57  & 49.42 & 51.10 & 49.89 & 50.90 \\
12          & 53.33 & 49.15  & 51.09 & 49.74 & 50.11 & 49.48 \\
13          & 53.33 & 49.15  & 48.17 & 50.83 & 50.01 & 49.99 \\
14          & 44.16 & 52.44  & 51.02 & 50.04 & 50.16 & 50.17 \\
15          & 57.50 & 52.97  & 49.56 & 50.00 & 50.11 & 49.88 \\
16          & 47.50 & 46.19  & 50.44 & 50.34 & 49.93 & 49.80 \\
17          & 54.17 & 51.37  & 49.34 & 49.31 & 49.89 & 49.81 \\
18          & 45.83 & 50.11  & 50.95 & 49.97 & 50.13 & 50.18 \\
19          & 46.67 & 50.58  & 50.88 & 50.07 & 49.83 & 50.09 \\
20          & 50.83 & 51.17  & 49.49 & 49.66 & 49.98 & 50.10 \\
\midrule
Average (\%)& 55.00 & 52.07  & 50.11 & 50.20 & 50.08 & 50.06 \\
\bottomrule
\end{tabular}
}
\end{table}

This is our first analysis of the Avalanche Effect, where we analyzed several TXT files. We flipped 0 to 20 bit positions for each file to check the avalanche effect. For flipping each bit position, we noted the percentage value of the avalanche effect. The average for all the files that underwent the Avalanche Effect test is above or equal to the ideal avalanche effect of 50\%. This phenomenal result ensures that HELO accomplishes the expected benchmark (Table \ref{tab12}).

\begin{table}[htbp]
\caption{Avalanche Effect Analysis of CSV Files}
\label{tab14}
\scriptsize 
\centering
\scalebox{1.25}{
\begin{tabular}{@{}lllllll@{}}
\toprule
Bit position & 1 MB& 6 MB& 10 MB& 67 MB& 84 MB&124 MB \\
\midrule
0            & 50.40 & 50.60 & 50.00 & 50.18 & 50.27 & 50.41 \\
1            & 49.48 & 49.62 & 50.00 & 49.35 & 50.28 & 49.40 \\
2            & 49.62 & 50.26 & 50.01 & 51.11 & 49.74 & 49.96 \\
3            & 49.36 & 50.60 & 50.00 & 50.90 & 49.17 & 50.04 \\
4            & 49.78 & 50.06 & 49.99 & 50.08 & 49.92 & 50.33 \\
5            & 49.81 & 51.09 & 50.00 & 49.64 & 51.03 & 49.78 \\
6            & 49.72 & 48.84 & 49.99 & 50.49 & 50.62 & 50.71 \\
7            & 49.01 & 49.81 & 50.00 & 50.49 & 49.68 & 48.73 \\
8            & 50.20 & 50.34 & 49.99 & 49.98 & 49.55 & 49.55 \\
9            & 50.45 & 49.46 & 50.00 & 50.07 & 50.69 & 49.87 \\
10           & 49.46 & 50.17 & 50.00 & 50.64 & 50.08 & 49.51 \\
11           & 49.23 & 50.37 & 50.00 & 50.38 & 49.23 & 49.59 \\
12           & 49.41 & 50.28 & 50.00 & 49.78 & 49.50 & 50.76 \\
13           & 50.53 & 50.32 & 50.00 & 50.06 & 50.32 & 49.80 \\
14           & 49.96 & 50.64 & 50.01 & 50.44 & 50.09 & 49.63 \\
15           & 49.66 & 49.89 & 50.00 & 50.09 & 50.11 & 50.20 \\
16           & 49.35 & 49.96 & 50.01 & 50.11 & 49.64 & 49.84 \\
17           & 50.19 & 49.75 & 50.00 & 50.75 & 50.43 & 51.12 \\
18           & 49.49 & 48.93 & 50.00 & 49.99 & 49.79 & 50.60 \\
19           & 50.33 & 49.50 & 50.00 & 51.08 & 50.07 & 49.73 \\
20           & 49.94 & 50.47 & 50.01 & 50.01 & 51.10 & 49.89 \\
\midrule
Average (\%) & 50.17 & 50.54 & 50.01 & 50.09 & 50.69 & 50.15 \\
\bottomrule
\end{tabular}
}
\end{table}

Our second analysis was to put CSV (Comma Separated Value) files on a test. We followed the same process of flipping bit positions and finding out the avalanche effect value for each file. The average is taken by finding the sum of all the bit position values and then dividing by the number of bit positions for each file. Therefore, the result is above 50\% for all files taken to test (Table \ref{tab14}).

\begin{table}[htbp]
\caption{Avalanche Effect Analysis of Image Files}
\label{tab15}%
\scriptsize 
\centering 
\scalebox{1.25}{
\begin{tabular}{@{}lllllll@{}}
\toprule
Bit position & 5 MB  & 20 MB  & 40 MB  & 80 MB  & 95 MB  \\
\midrule
0            & 50.13 & 50.15 & 50.05 & 50.22 & 50.10 \\
1            & 50.12 & 50.03 & 49.97 & 50.10 & 50.04 \\
2            & 49.83 & 49.96 & 50.01 & 49.87 & 49.98 \\
3            & 49.85 & 50.06 & 49.96 & 49.89 & 49.95 \\
4            & 49.97 & 50.05 & 50.02 & 49.94 & 50.04 \\
5            & 50.05 & 50.13 & 49.98 & 49.93 & 49.93 \\
6            & 49.96 & 49.98 & 50.04 & 50.02 & 50.02 \\
7            & 49.94 & 49.91 & 49.98 & 49.87 & 49.99 \\
8            & 49.83 & 49.89 & 49.94 & 50.17 & 49.98 \\
9            & 49.75 & 50.22 & 49.90 & 49.79 & 50.07 \\
10           & 49.91 & 50.04 & 49.87 & 50.21 & 49.98 \\
11           & 49.79 & 50.02 & 50.00 & 49.68 & 50.10 \\
12           & 50.09 & 50.04 & 49.95 & 49.97 & 49.96 \\
13           & 50.23 & 50.07 & 49.98 & 50.02 & 49.92 \\
14           & 50.31 & 49.91 & 50.01 & 50.11 & 49.90 \\
15           & 50.06 & 49.82 & 49.97 & 50.14 & 50.12 \\
16           & 49.60 & 50.08 & 50.15 & 50.12 & 49.87 \\
17           & 50.26 & 50.05 & 49.96 & 50.10 & 49.83 \\
18           & 49.87 & 49.99 & 50.04 & 49.73 & 49.96 \\
19           & 49.77 & 50.00 & 49.95 & 49.83 & 49.88 \\
20           & 50.20 & 50.21 & 50.04 & 50.18 & 50.06 \\
\midrule
Average (\%) & 50.16 & 50.18 & 50.04 & 50.20 & 50.08 \\
\bottomrule
\end{tabular}
}
\end{table}

Since the image file needs some rendering and other process capabilities to run and view it, it must check whether the avalanche effect value decreases. Fortunately, after testing, we found that it gives an average value equal to or slightly more than the standard avalanche effect value, which is 50\%. This test provides us with an important analysis and evaluates if our algorithm is secure to use or not. Hence, we can conclude that this test has provided us with reliability and security assurance (Table \ref{tab15}).

\subsection{Justification for Choosing Specific Components in HELO}
Table \ref{tab41} summarizes the performance evaluation and justification for the cryptographic components selected in our algorithm. It outlines each component against established alternatives based on the analyses conducted, focusing on efficiency, resource consumption, and security benefits.

\begin{table}[htbp]
\centering
\caption{Justification of Selected Components Compared to Common Alternatives}
\label{tab41}
\resizebox{\textwidth}{5.2cm}{
\begin{tabular}{|p{3cm}|p{3.5cm}|p{2cm}|p{7cm}|}
\hline
\textbf{Component} & \textbf{Selected Algorithm} & \textbf{Alternative} & \textbf{Justification}\\
\hline 
Symmetric Encryption and Encryption Integrity  & ChaCha20-Poly1305 & AES, Blowfish, Fernet & ChaCha20-Poly1305 were chosen over AES and Fernet due to their superior performance in software-based environments, requiring fewer CPU cycles and memory resources. Its constant-time execution model provides strong resistance to timing attacks, a critical consideration for IoT security. Additionally, its AEAD construction allows simultaneous encryption and authentication in a single pass, reducing runtime and energy consumption compared to dual-pass schemes like Fernet, which is discussed in CPU processing (~\ref{cpupro}) as well as in the runtime analysis (~\ref{runtime}). \\
\hline 
Hash Function & SHA3-256 & SHA2-256, SHA-1 & Sponge-based design offers better resistance to length extension and collision attacks; suitable for future post-quantum considerations \cite{Bertoni2010}.\\
\hline 
Key Exchange & ECDH  & RSA, DH & Smaller key sizes, lower computational cost, and better efficiency in low-power embedded systems that provide forward secrecy \cite{adarbah2023security}. \\
\hline 
Digital Signature & ECDSA & RSA & Shorter keys and signatures reduce transmission and storage overhead, and faster signature generation/verification is better suited for IoT and P2P communication \cite{farooq2019ecdsa}. \\
\hline
Nonce & Unique per session, counter-based & Random Nonce & Ensures cryptographic freshness; prevents nonce reuse vulnerabilities in stream cipher modes; lightweight implementation for deterministic nonce generation.\\
\hline
\end{tabular}
}
\end{table}

\section{Conclusion}\label{sec7}
The HELO cryptographic system addresses numerous security issues identified during the investigation by implementing a comprehensive and secure system that ensures that no vulnerabilities are found in many existing models. The system can handle security issues related to authorization, authentication, or even data access very effectively. Furthermore, this system goes about a unique way of introducing a new key exchange algorithm to ensure that the data and system itself are fully secured. The research demonstrates a multi-layered security approach that provides detailed data and access control layer security, which can protect data against several types of breaches and cyberattacks. These security measures reduce the risk of several types of attacks, which ensures that precious data is secured and also prevents hackers from exploiting the weakness in the system or compromised end-user credentials. However, the algorithm didn’t reduce its performance while ensuring strong protection. It performs faster than the existing algorithms in many cases. Hence, this HELO cryptographic system has been proven to be one of the best approaches for data security as lightweight cryptography in IoT devices.

\section{Future Work}\label{sec8}
In the rapidly evolving landscape of computerized security, the planning and ongoing upgrades of cryptographic frameworks play an urgent role in defending sensitive data and guaranteeing secure communication over different technological domains. Here, HELO is a modern cryptographic approach that offers robust security against a wide range of threats. Therefore, ongoing research and development are crucial to address emerging vulnerabilities and opportunities to identify for system enhancement. One of the key focuses of future HELO studies is improving resilience against critical cyberattacks, which involves empowering detection and response capabilities. With that, strengthening encryption mechanisms and introducing adaptive defenses can enhance the dynamic response to new threat models. In addition, the focus on the future study of blockchain and distributed laser technologies in IoT security is important to us as it offers transparency, irreversibility, and decentralized trust \cite{madhi2023iot}. Another important area of our future study is the examination of homomorphic encryption, which allows calculations on encrypted data without revealing the underlying information. Parallel to technological development, we will discover regulatory, legal ideas, and security protocols in our future work to follow the International Data Protection Act and cybersecurity policy so that the cryptographic applications are safe, valid, and trustworthy \cite{farhan2009security}, \cite{ali2018security}. However, the current configuration of the system introduces some limitations. Particularly, the absence of real IoT hardware limits the actual generality of the results in the embedded environment. To overcome this, our future work will involve deploying the system on real IoT devices and evaluating performance, energy consumption, scalability, and computational overhead under more realistic conditions. By addressing both existing obstacles and future possibilities, our comprehensive goal is to strengthen digital trust and fundamental pillars of privacy in the era of the Internet.

\end{document}